 \documentclass  [twocolumn]{aa}     % for a paper on 2 column
\usepackage{graphicx}
\usepackage{txfonts}
\usepackage[dvips]{color}
\usepackage[normalem]{ulem}

\begin {document}
 \title{The structure of solar radio noise storms}

   \subtitle{}

   \author    {C. Mercier \inst{1} \fnmsep \thanks{ }     %  le "\fnmsep \thanks{ }" a un effet nul
         \and   P. Subramanian \inst{2} \fnmsep \thanks{ }
         \and   G. Chambe \inst{1} \fnmsep \thanks{ }
         \and   P. Janardhan \inst{3} \fnmsep \thanks{ }
                                   }

   \institute  {LESIA-Observatoire de Paris, CNRS, UPMC, Univ. Paris-Diderot %\\
                   % le retour a la ligne semble automatique pour les instituts, mais pas pour les emails
          \and   Indian Institute of Science Education and Research, Sal Trinity Building, 
                         Pashan, Pune -411021, India % \\
          \and   Astronomy and Astrophysics Division, Physical Research Laboratory
                        Ahmedabad - 380 009 % \\
                          \email  {claude.mercier@obspm.fr}\\                    % le \\ force retour a la ligne
                          \email  {p.subramanian@iiserpune.ac.in}\\         %     pour le email suivant.
                          \email  {gilbert.chambe@obspm.fr}\\
                          \email  {jerry@prl.res.in}\\
                         }                            % fin du "institute"     

   \date{Received xxx, 2012;  Accepted xxx, 2022}

% \abstract{}{}{}{}{} 
% 5 {} token are mandatory
%---------------------------------------------  pattern of the line length ----- -----------------------------------------------

\abstract
{The Nan\c{c}ay Radioheliograph (NRH) routinely produces snapshot images of the full sun (field of view   $\sim 3 R_{\odot}$) at 6 or 10 frequencies between 150 and 450 MHz, with typical resolution 3 arcmin and time cadence 0.2 s.  Combining visibilities from the NRH and from the Giant Meterwave Radio Telescope (GMRT) allows us to produce images of the sun at 236 or 327 MHz, with the same field as the NRH, a resolution as low as 20 arcsec, and a time cadence 2 s.}
{We seek to investigate the structure of noise storms (the most common non-thermal solar radio emission) which is yet poorly known. We focus on the relation of position and altitude of noise storms with the observing frequency and on the lower limit of their sizes.}
{We  use an improved version of a previously used method for combining NRH and GMRT visibilities to get high-resolution composite images and to investigate the fine structure of noise storms.  We also use the NRH data over several consecutive days around the common observation days to derive the altitude of storms at different frequencies.}
{We present results for noise storms on four days.  The results consist of an extended halo and of one or several compact cores with relative intensity changing over a few seconds.  We found that core sizes can be almost stable  over one hour, with a minimum in the range 31-35 arcsec (less than previously reported) and can be stable over one hour.  The heliocentric distances of noise storms are $\sim 1.20$ and 1.35 $R_{\odot}$ at 432 and 150 MHz, respectively.  Regions where storms originate  are thus much denser than the ambient corona and their vertical extent is found to be less than expected from hydrostatic equilibrium.}
% conclusions heading (mandatory)
{The smallest observed sizes impose upper limits on broadening effects due to scattering on density inhomogeneities in the low and medium corona and constrain the level of density turbulence in the solar corona.  It is possible that scatter broadening has been overestimated in the past, and that the observed sizes cannot only be attributed to scattering.  The vertical structure of the noise storms is difficult to reconcile with the classical columnar model.}

%  La commande des couleurs marche de facon tres capricieuse dans l'abstract.  Il suffit de deplacer l'accolade de fin d'un seul caractere pour que tout devienne rouge.  Donc on supprime les ordres de couleur dans l'abstract.  
  

\maketitle

% \end {document}

%---------------------------------------------  pattern of the line length ----- -----------------------------------------------

 									% spaces :    \ ,   \ :    \ ;    (without blanks after \)
									%  {\color {red}  text}
									%  \txtbf {text en gras}

\section{Introduction}
Type I noise storms are the most common non-thermal solar radio emission in the decimetric-metric range.  Noise storms also exist in the decametric range (below 100 MHz) and exhibit different characteristics.  These storms are usually referred to as decameter noise storms and are not dealt with here.  Type I noise storms (merely referred to as noise storms from now on) can last for hours to days and reveal long-lived production of suprathermal electrons not directly related to flares (Le Squeren, \cite{lesqueren}).  
Extensive descriptions have been given by Elgar\o y (\cite{elgaroy}) and Kai \emph{et al} (\cite{kai}).  They consist of a broadband continuum ($\Delta f / f \sim 1$) with superimposed Type I bursts, more frequent below 250 MHz.  Bursts have durations $\le 0.5 \: sec$ and bandpass $\Delta f / f \sim 3 \%$.  They occur in chains of some tens of seconds with typical intervals 1-2 sec., drifting slowly toward low frequencies.  Continuum and bursts are often highly circularly polarized in the o-mode of the underlying photospheric magnetic field (eg. Mercier \emph{et al}, \cite{cons}, Kai \emph{et al}, \cite{kai}).  The apparent size of noise storms is $\le 3$ arcmin.  The apparent brightness temperature $T_b$ of the continuum can exceed $10^{8}$ K (Kai \emph{et al}, \cite{kai}) and even can reach $\sim10^{10}$ K (Kerdraon and Mercier, \cite{ker-mer-AA}).

The consensus is that radio emission occurs at the fundamental plasma frequency and is due to suprathermal electrons trapped in closed flux tubes.  This explains the strong polarization in the o-mode and the high values for $T_b$.  Theories have been proposed by Melrose (1980), Benz and Wentzel (1981) and Spicer \emph{et al} (1982), involving the coalescence of plasma waves with low-frequency waves (ion-acoustic or lower-hybrid).  Melrose did not specify the origin of fast electrons but Benz and Wentzel and Spicer ascribed their production to reconnection or to weak shocks associated with newly emerging flux. Subramanian \& Becker (\cite{subra-beck-1},  \cite{subra-beck-2}) considered a generic second-order Fermi acceleration mechanism for generating these electrons. They found an overall efficiency $\sim$ $10^{-6}$ for the emission process.

Only few instruments have produced one-dimensional (1D) or two-dimensional (2D)  images of noise storms : the Culgoora radioheliograph (CRH, 2D, first at 80 MHz, later also at 160, 320, and 43 MHz, Labrum, 1985), the Nan\c{c}ay radioheliograph (NRH), first 1D at 169 MHz, later 2D at 5, 6, or 10 frequencies from 150 to 450 MHz, Kerdraon \& Delouis, 1997), the Very Large Array (VLA, 2D at 333 MHz, see eg. $http://www.vla.nrao.edu$) and the Giant Meterwave Radio Telescope (GMRT, Ananthakrishnan \& Rao, 2002) combined with the NRH at 327 MHz.  The CRH (closed in 1986) and the NRH are dedicated to the sun, and the VLA and the GMRT only occasionally observe it. 

In principle, imaging noise storms at different altitudes requires : $\:$ i) a field of view wider than the sun since several storms often coexist, $\:$ ii) a resolution better than their typical apparent size to see a possible fine structure, and $\:$ iii) simultaneous observations at several frequencies over their bandpass.  Until now, these conditions were never fulfilled at the same time.  This explains why the structure of noise storms is still poorly known 

The CRH, the NRH, and the VLA in its compact C configuration (VLA-C) have wide fields of view, but their resolution ($\sim$ 1 arcmin) is hardly less than the storm size.  No internal structure in storms can be observed.  Kai \emph{et al} (\cite{kai}) reported that bursts and continuum have approximately the same location and that the apparent size increases with decreasing frequency.  This was confirmed by Malik and Mercier (\cite{malik}) from NRH observations at 164, 236, 327, and 410 MHz.  Based on analysis of data covering seven days, they also reported that the observed positions were close to each other from 236 to 410 MHz, with small changes roughly parallel at the different frequencies, and that the positions of the bursts and continua widely overlap, the bursts being slightly narrower.  In some cases apparent sizes are observed down to the resolution.  Kerdraon (\cite{ker1973}) uses the NRH at 169 MHz, reported sizes of 1.3 arcmin, and Habbal \emph{et al} (\cite{habbal}), from observations on two days with the VLA-C at 333 MHz, reported sizes of $57 \times  47$ arcsec.  Storms with both LH and RH circular polarizations were occasionally reported, essentially from early observations with the CRH and from some observations with the VLA.  However, from a careful analysis of these reported cases and from an extensive analysis of NRH data, Malik and Mercier (\cite{malik}) concluded that there was no evidence of bipolar structure and that most of the reported cases should be either artifacts or (in some cases only) pairs of separate noise storms.

High-resolution ($<$ 10 arcsec) observations with the VLA at 327 MHz in the extended A-configuration revealed some internal fine structure.  Lang and Willson (\cite{lang-will}) described one storm as consisting of four compact sources each about 40 arcsec in angular diameter, arranged within an elongated 40 $\times$ 200 arcsec source".  However, the time resolution was only 13 and 30 s and the dynamical range in images was limited by the small number (12) of the antennas used.  Kerdraon \emph{et al} (\cite {ker-lang}) presented images of two bursts occuring at different times during a storm, with typical sizes of 45 arcsec and positions separated by $\sim$ 50 arcsec, one being polarized and the other unpolarized.  Zlobec \emph{et al} (\cite{zlobec}) attempted to derive the smallest spatial scales involved in noise storms.  From observations on two days at 333 MHz, they did not detect any significant power for baselines $>5000 \: \lambda$,  and gave minimum reliable sizes of $\sim 40$ arcsec, the smaller derived sizes being considered as questionable because of the poor $uv$-coverage (not all antennas were available).  Mercier \emph{et al} (\cite{mercier2006})  combined complex visibilities from the NRH and the GMRT at 327 MHz for one day.  The resulting images had potentially the same resolution as with the VLA but had a larger field of view.  Three noise storms were present, and Mercier \emph{et al} gave results for the most intense storm.  Because of the difficult intercalibration between both instruments, the accepted baselines were limited to about 6000 $\lambda$. Sizes of $\sim 50$ arcsec were reported, in general agreement with the results of Zlobec  \emph{et al}, (\cite{zlobec}).

Some of these results, namely the increase of the observed sizes with decreasing frequencies, are consistent with the classical columnar model (Kai \emph{et al}, \cite{kai}) in which emission at different frequencies originates at different altitudes in the same flux tube.  However, Lang and Willson (\cite {lang-will}) pointed out that the observed complexity in the structure of noise storms may well rule out this simple model.  Malik and Mercier (\cite {malik}) also pointed out that the close positions observed at different frequencies could be difficult to explain with a flux tube in hydrostatic equilibrium.  In fact, very little is known concerning the vertical extent of noise storms.  In the absence of stereoscopic observations, the only possibility is to derive the altitudes of persistent storms at several frequencies from their apparent rotation (assumed to be  rigid) during several consecutive days, but this has never been done. 

The highest $T_b$ values of the continua and the smallest sizes of the continua and bursts are of interest, : the highest values of  $T_b$ constrain the emission theories and the smallest sizes put an upper limit to the broadening resulting from the scattering of radio waves by  coronal density inhomogeneities (Bastian 1994, Subramanian \& Cairns 2011).  Up to now, the highest reported $T_b$ values are of the order of $10^{10} \: K$, in agreement with the maximum predicted from plasma emission models.  These values, however, were obtained with a relatively low resolution ($>1$ arcmin), and could have been underestimated. 

%---------------------------------------------  pattern of the line length ----- -----------------------------------------------

\begin{figure}[!ht]   %   rh0gm1,  rh1gm1   cas complexe  2006 avr 06
                                  % version 2 colonnes :  
                                  %   -  {figure*} pour Žtaler sur deux col.  (avec aussi   end{figure*}  ici inutile
                                  %  -  {figure}  pour utiliser une seule col (dim  max 4.45  pour y mettre les deux fig)
\centering
\includegraphics[width=4.45cm]{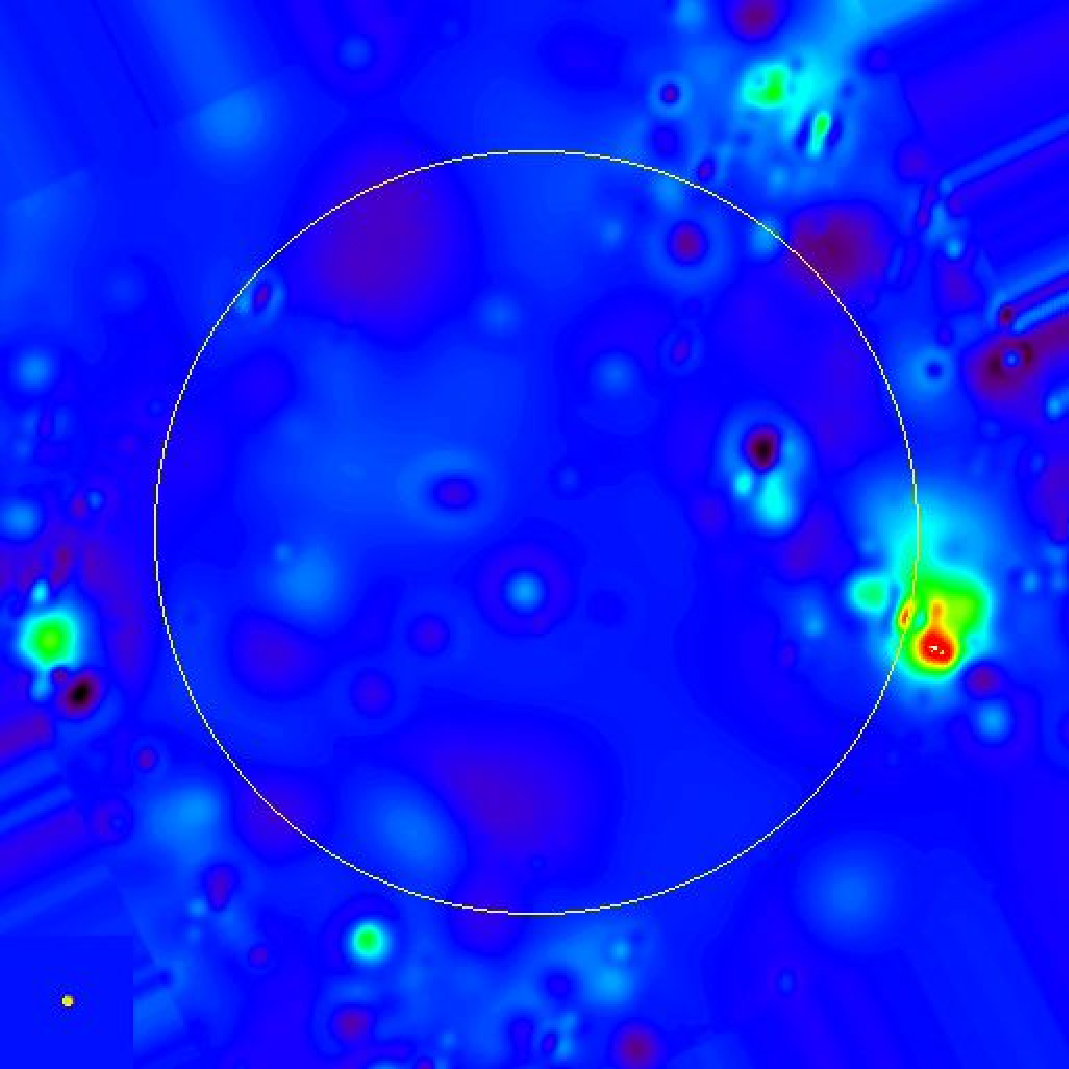}  % 4.55  pour version 1 col
\includegraphics[width=4.45cm]{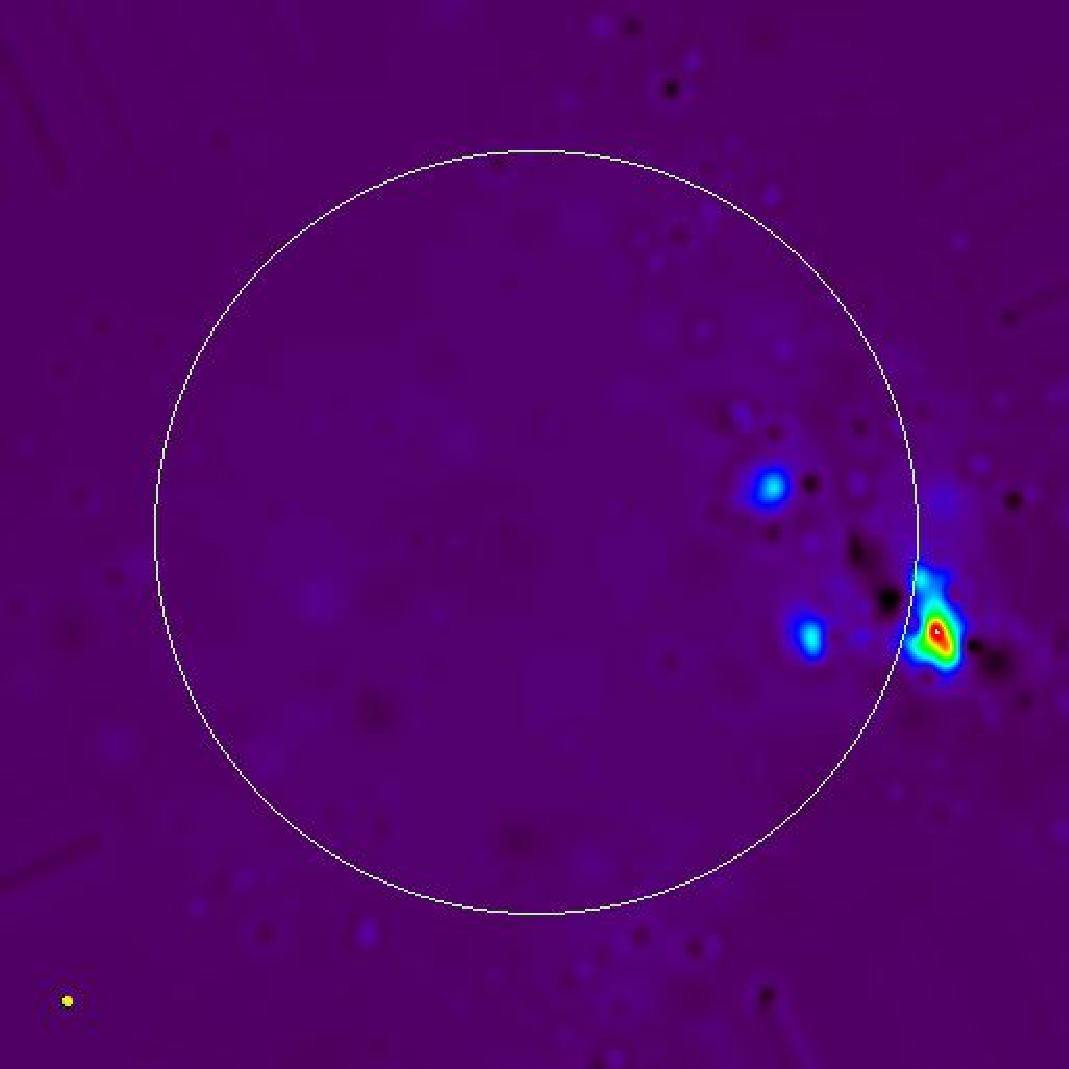}  % 4.45 ------------------ 2 col
\includegraphics[width=4.45cm]{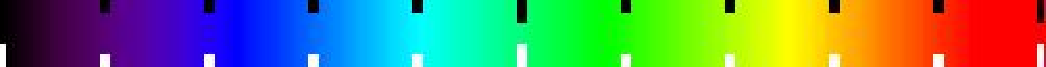}
\includegraphics[width=4.45cm]{fig_1cd.eps}
\caption{Examples of images with the GMRT alone (left) and with NRH+GMRT(right)  for a complex situation on Apr. 06, 2006 at 10:01:28 UT.  The intensity scale (shown at bottom) is linear from black (lowest level) to red (highest level).  The color of the background (zero level) results from the value of the deepest negative artifact. This color scale makes low-level negative artifacts more visible than the linear black and white scale. The resolution ($\sim20$ arcsec) is indicated at bottom left.  The circle is the optical limb.}
\label{aliasing}
\end{figure}

The above reviewed observations show that a spatial resolution substantially less than 1 arc min ($\sim20$ arcsec or less, corresponding to baselines up to 10000 $\lambda$) is needed to properly investigate the fine structure of noise storms.  Only very few such high-resolution observations with the VLA are available since its limited field of view restricted studies to a small number of intense bursts.  The advantage of combining the NRH and GMRT data, as first presented by Mercier \emph{et al} (\cite {mercier2006}) is to get both high resolution and a wide field of view, allowing them to use the common observations entirely.  In this paper, we present results from simultaneous observations with the NRH (150, 164, 236, 327, 410 MHz) and GMRT (236 or 327 MHz).  These frequencies cover most of the frequency range of noise storms.  We give results for four noise storms observed on different days, with an improved NRH$/$GMRT intercalibration, allowing a better resolution than in Mercier \emph{et al (\cite {mercier2006})}.   Two cases are complex, with up to three other simultaneous noise storms, and could not have been investigated without combining the two instruments.  This is illustrated in Fig. \ref {aliasing}.  We also present an analysis of the positions of the storms at all the NRH frequencies over several consecutive days before and after the common observations to investigate their vertical extent.

%---------------------------------------------  pattern of the line length ----- -----------------------------------------------

\section {Observations, data selection, and processing}
We used all available joint GMRT / NRH observations of noise storms (table 1).  Joint observations are possible over $\sim$ 08:30-12:00 UT, but those presented here are limited to $\sim$ 1hour.  The storm on Aug. 27, 2007, already presented by Mercier \emph{et al} (\cite {mercier2006}), is used again since we improved the data processing.  The NRH uses an integration time $\tau =$ 0.125 s, shorter than typical burst durations, whereas the GMRT uses a longer $\tau$ of 2.1 or 17 s.  The NRH data were integrated over the time intervals used by the GMRT.  The final time cadence is thus that of the GMRT.  The time profiles of bursts are therefore smoothed and their maximum intensity might be underestimated.

% \end {document}
%\begin {table} [h]  \begin {center}                                   [!ht]
\begin {table} [!ht]  \begin {center}  %{table*} pour toute la largeur de la page. Il faut   \end {table*}
                                                             %                 Ici inutile car la table tient (juste) sur une colonne.
                                                              %  Il faut  [!ht]  et non  [h], sinon la table est reportee a la fin;
\title {Table 1 : Noise storms observed with GMRT and NRH.}  \\ [1.5ex]     % [1.5ex] 
\begin{tabular}[c]{lccccccccc} \hline\hline   % {llllr}
%date                       &  GMRT &  GMRT  &  & \multicolumn {6} {c}   ---------------------- NRH %---------------------- \\ %  [0.5ex]

date                &     GMRT    & \multicolumn {6} {l} -------------------- NRH --------------------- \\ %  [0.5ex]

                        & $\tau$        & \multicolumn {6} {l}  ----------------------- $f$ ---------------------- \\
%                              &  MHz &  sec  &   \\
\hline
2002 Aug 27 & 17  &          & 164  &  236   &  327*  &  410  & 432 \\
2003 Jul 15   & 2.1 & 150  & 164  &  236*  &  327   &  410  & 432 \\   %[1.0ex]
2004 Aug 14 & 2.1 & 150  & 164  &  236   &  327*  &  410  & 432 \\   %[1.0ex]
2006 Apr 06  & 2.1 & 150  & 164  &  236*  &  327   &  410  & 432 \\   % [1.0ex]
\hline
\multicolumn {8} {l} {\small $\tau$ integration time (s) for GMRT, $f$ frequency for NRH (MHz).} \\
\multicolumn {8} {l} {\small The common frequency with GMRT is indicated with *.} \\
\end{tabular}
\end {center}
\end {table}
%                                                                  {\color {red}  text}
%---------------------------------------------  pattern of the line length ----- -----------------------------------------------
\subsection {Calibration of each instrument}
The calibration of the NRH is done with Cygnus A about every two weeks and is completed with a self-calibration.
% The NRH is calibrated with Cygnus A about every two weeks.  A self-calibration makes the instrumental phase correction more accurate.  
Since all possible baselines are not achieved, a classical procedure based on phase closure cannot be used and a specific method was developed.  This method relies on the fact that the NRH $uv$-coverage is both dense and regular around the origin.  For a perfectly phased array, the image of a point source is a sinc-like function (at 327 MHz, the beamwidth is $\sim2$ arcmin  and the field of view is $\sim1$ deg.).  With imperfect calibration, side lobes are less regular and larger, especially far from the central beam.  Antenna phases are obtained though an iterative reduction of the zero and second order momenta of side lobes.  Intense and compact bursts, producing a limited smoothing of side lobes, are used as calibrators.  Using baseline redundancies reduces the number of free phases from 48 down to 17, making the procedure more rapid and robust.  The accuracy in phase is $<$ 5-10 deg. and the artifacts on the NRH clean images are reduced  to $\sim$ 5\% and 1-2\% for peak and rms values, respectively. 

A typical solar observation sequence with the GMRT comprises around 30 minutes on the Sun, with 30 dB attenuators inserted, followed by around five minutes on a nearby cosmic phase calibrator, with the attenuators removed.  This sequence is repeated for the duration of the observation.  The values of the 30 dB attenuators are uncertain by around 10\% (0.4 dB). The automatic gain control mechanism is switched off throughout the observation.  We solve for the antenna gains using data from the phase calibrator, which are then applied to the solar data.  The rms uncertainty on the amplitude of antenna + attenuator gains is around 15 \%.  An absolute flux calibration is achieved a-posteriori by comparing with NRH visibilities (section 2.3).

\subsection {Phase intercalibration}
It should be ensured, before combining visibilities,  that the storm positions derived from both instruments coincide within less than the expected resolution of a few arcsec.  Apparent positions may differ for two reasons:  i) both instruments may have different systematic positional errors (up to $\sim 20$ arcsec for the NRH) and,  ii) ionospheric effects are different for each instrument.  They can be $\sim$ 1 arcmin at 236 or 327 MHz.  We directly measured the position differences $\Delta X$ and $\Delta Y$ along the EW and NS directions.  A factor $exp(-i2\pi(u\Delta X + v\Delta Y))$ was then applied to the GMRT complex visibilities before  combining them with the NRH visibilities.
 
The position differences $\Delta X$ and $\Delta Y$ could in principle be derived from the phase differences between the NRH and GMRT complex visibilities where the $uv$-coverages coincide, by fitting the phase differences to a linear function of the spatial frequencies $u$ and $v$.  However, there are generally no exact coincidences and even "approximate" coincidences are rare, in spite of the NRH $uv$-coverage density.  Indeed the distance between the NRH and GMRT points should be much less than the variation scale  $\sim 1/L$ of the complex visibility, where $L$ is the typical mutual distance between solar radio sources.  Moreover these phases are affected by noise and interference and the procedure would only partially use information from both instruments.  It follows that $\Delta X$ and $\Delta Y$ would be uncertain.  

\begin{figure}[!ht]   %  version 2 colonnes : 
                                    %  - {figure*}  pour  Žtaler sur deux colonnes    ici inutile
                                    %  - taille max 9.1 cm max pour tenir sur une seule colonne
\centering
\includegraphics[width=9.1 cm, height = 3.2 cm]{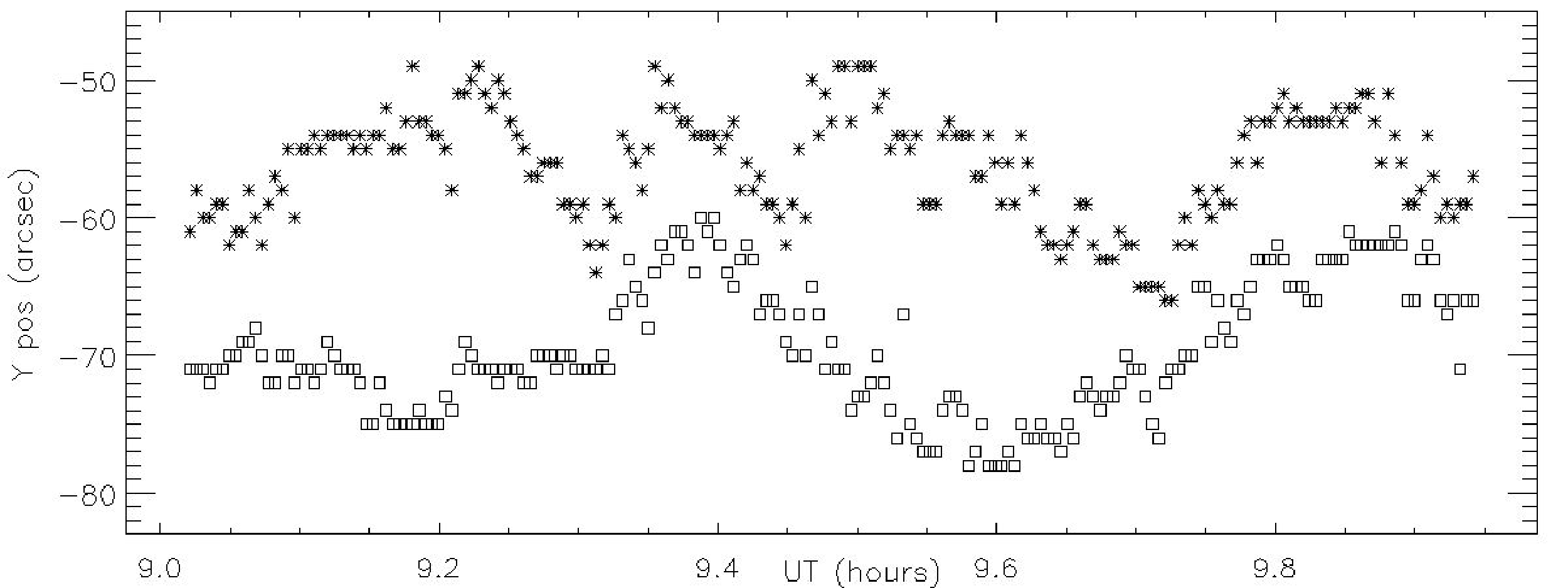}           % 3.2
\includegraphics[width=9.1 cm, height = 2.5 cm]{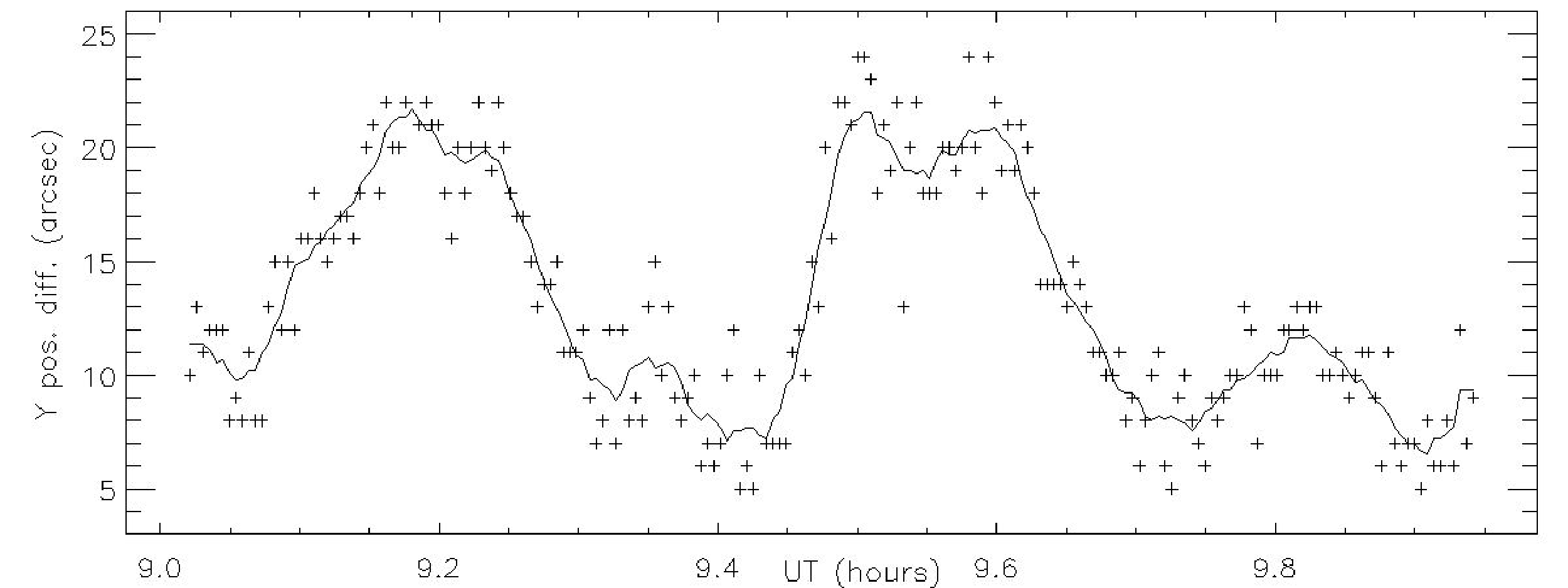}
\caption{Top: NS positions (arcsec) versus time (UT) for the main noise storm on Aug. 27, 2002 for NRH (squares) and GMRT (asterisks).  Bottom: difference  between GMRT and NRH NS positions (arcsec).  The continuous curve gives the sliding average over two minutes.}
\label{diffpos 2002aug27}
\end{figure}

Instead, we preferred to directly measure the positions in images separately obtained from the NRH and the GMRT, which more completely uses information.   For this, we limited the GMRT resolution to the NRH resolution, accepting only GMRT baselines up to $\sim 3$ km.  Direct measurement of the position of maximum in both images would be affected by rounding errors since images are calculated on grids with a finite step ($\sim 20 $ arcsec in this case).  Instead, we measured the barycenter position of the brightness distribution where it is larger than a given fraction $m$ of its maximum.  A value $m=0.8$ gave good results, ensuring also that the obtained positions are not sensitive to possible extended patterns at low levels.  Fig. \ref{diffpos 2002aug27} shows an example: NRH and GMRT north-south positions show partially correlated variations with time (top), the correlated part being mainly of solar origin.  Their difference (bottom) involve mostly timescales $\sim 25$ minutes, typical of transient ionospheric disturbances.  The accuracy, estimated from the dispersion of points around the solid line, is $<5$arsec.  We obtained similar results in the EW and NS directions for all days.  Given the aspect of the results in Fig.  \ref{diffpos 2002aug27}, we adopt a sliding average over two minutes for $\Delta X$ and $\Delta Y$.  

\subsection {Amplitude intercalibration}
The amplitude intercalibration results from comparing amplitudes of the NRH and GMRT complex visibilities in the region of good overlap.  This occurs for baselines $<1000$ m, where the NRH $uv$-coverage is dense, excluding baselines  $<150$ m, which are lacking in the GMRT.  For each data set (i.e., for each snapshot comprising one integration time $\tau$),  a multiplying factor $C$ (applied to the GMRT visibilities) was determined in such a way that the amplitudes of the NRH and GMRT complex visibilities, considered as functions of only the distance to the origin, coincide at best in the overlapping domain.  Ideally, $C$ should remain stable over the common observation duration.  This was the case for two days (Jul. 15, 2003 and Aug. 14, 2004).  For the two other days, $C$ was slowly drifting by up to 20\%.  In order to reduce possible gain variation effects, we used a sliding average over two minutes. 

\subsection {Image calculation}
After intercalibration, we obtained snapshot images through the usual procedure: a Fourier transform of the whole set of data, followed by a deconvolution.  We used the so-called clean procedure, improved with a scale analysis, as described in Mercier \textit{et al} (\cite{mercier2006}).   The efficiency of the intercalibrations was then checked by comparing the results with those obtained by applying to the GMRT visibilities an extra position shift and/or an extra amplitude coefficient.  Results were globally better without such extra corrections.  The sensitivity was $\sim5$ arcsec for position and 10\% for amplitude.  Images were better than in Mercier  \emph{et al} (\cite{mercier2006}).  The addition of longer baselines resulted in a better resolution, down to 15 arcsec, taking the final tapering into account.
  
%\end {document}                     %  {\color {red}  text}

%---------------------------------------------  pattern of the line length ----- -----------------------------------------------

\begin{figure}[!ht]     %   exemple 2002 aug 27
                                     %  {figure*}  et end{figure*}  pour etaler sur toute la largeur.  Ici inutile.
                                      % width=3.38cm  pour version 1 colonne (max pour 4 figures sur une ligne)
                                      % width=A VOIR  pour version 2 colonnes                                                       
\centering
%  images rainbow avec barres
    \includegraphics[width=4.45cm]{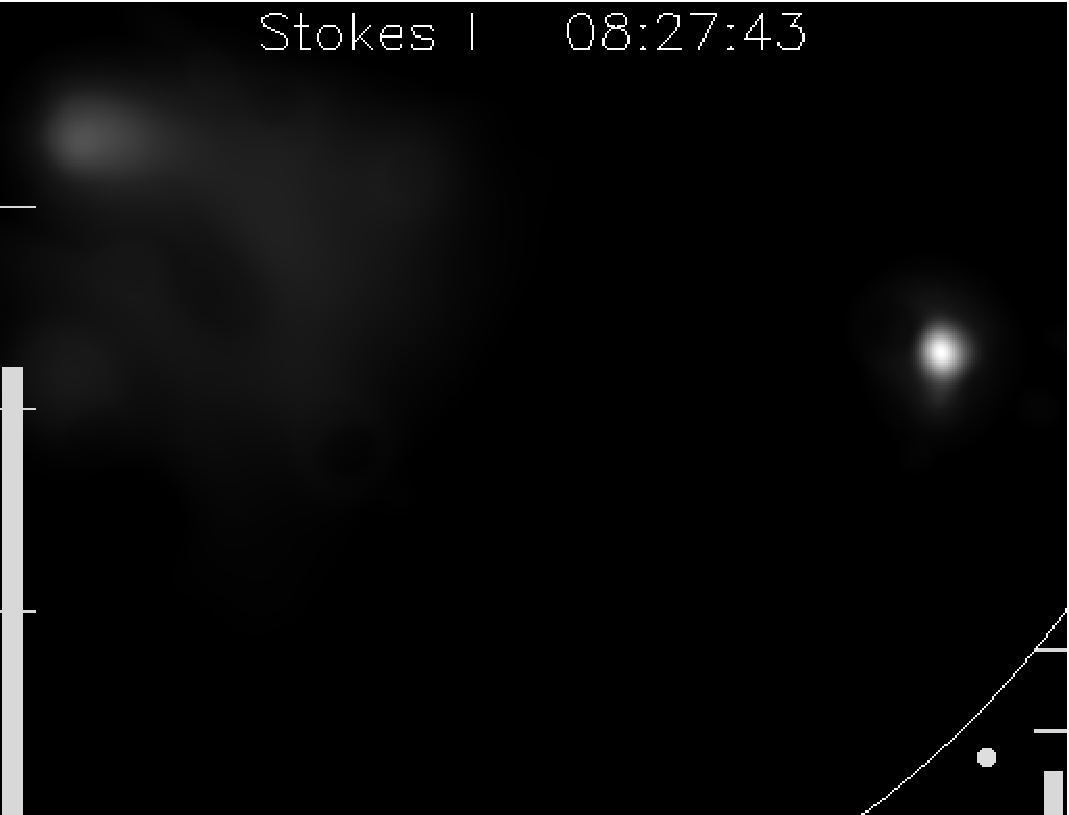}    % 08/27/43 plante
    \includegraphics[width=4.45cm]{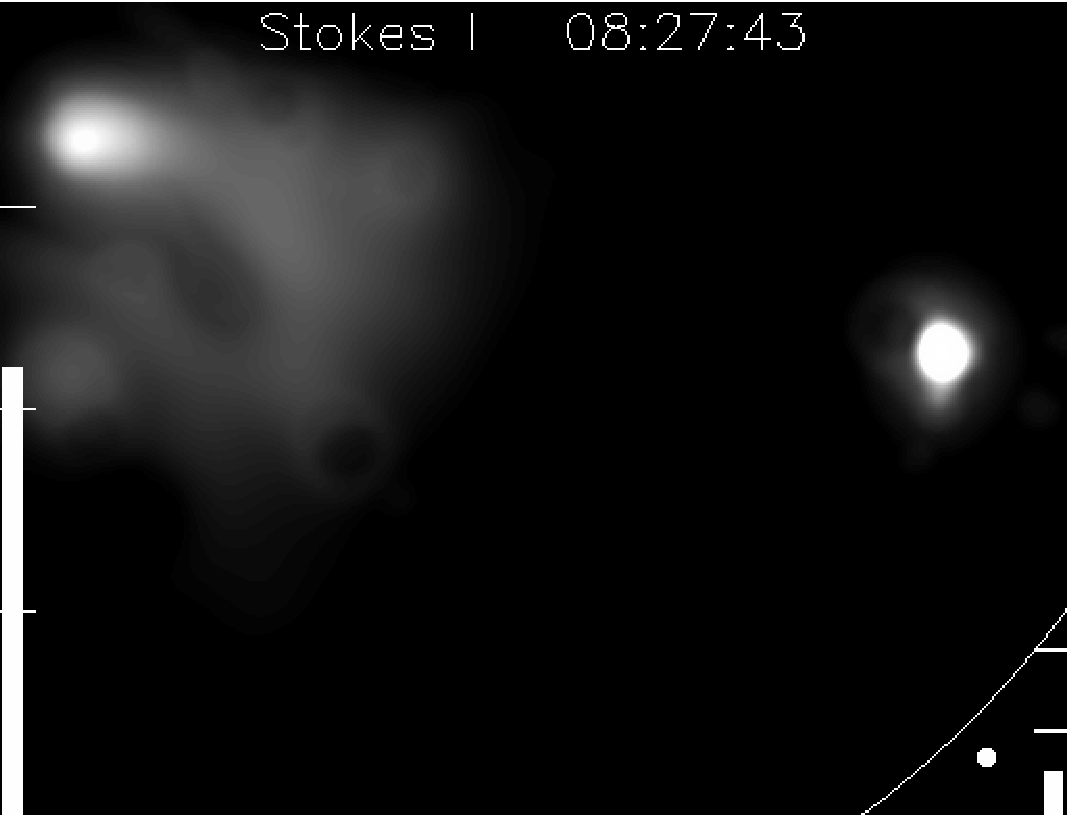}   %  4.55 version 1 col
\caption{Composite images for Stokes I at 327 MHz on Aug. 27, 2002 at 08:27:43.  The color scales are linear BW.  In the right frame, the image is saturated at 32\% of its maximum.  The limits of the field of view (units of $R_{\odot}$) is ($-0.25, +0.80)$ for EW and $(-0.80, 0.00)$ for NS.  The solid line is the solar limb. The storm near the western limb is simple.  The storms near the center of the disk are more extended.  The aspect of the main storm in Stokes V (not shown) is similar to that in Stokes I.  The vertical bars on the left side of images are proportional to $log(T_{B \: max}$), with $T_{B \: max}$ ranging from $10^6$ K at the bottom of the frame, to $10^{10}$ K at its top.  Negative artifacts have been suppressed for giving a dark background at zero level.  The relative value of the deepest negative artifact is proportional to the length of the vertical bars on the right side, with ticks at 10\% and 20\%. Same for Fig. 4 and 5.}

\label{images 2002aug27}
\end{figure}
\begin{figure}[!ht]     %   exemples 2003 jul 15 et 2004 aug 14 regroupes sur une seule figure.
\centering
\includegraphics[width=4.45 cm]{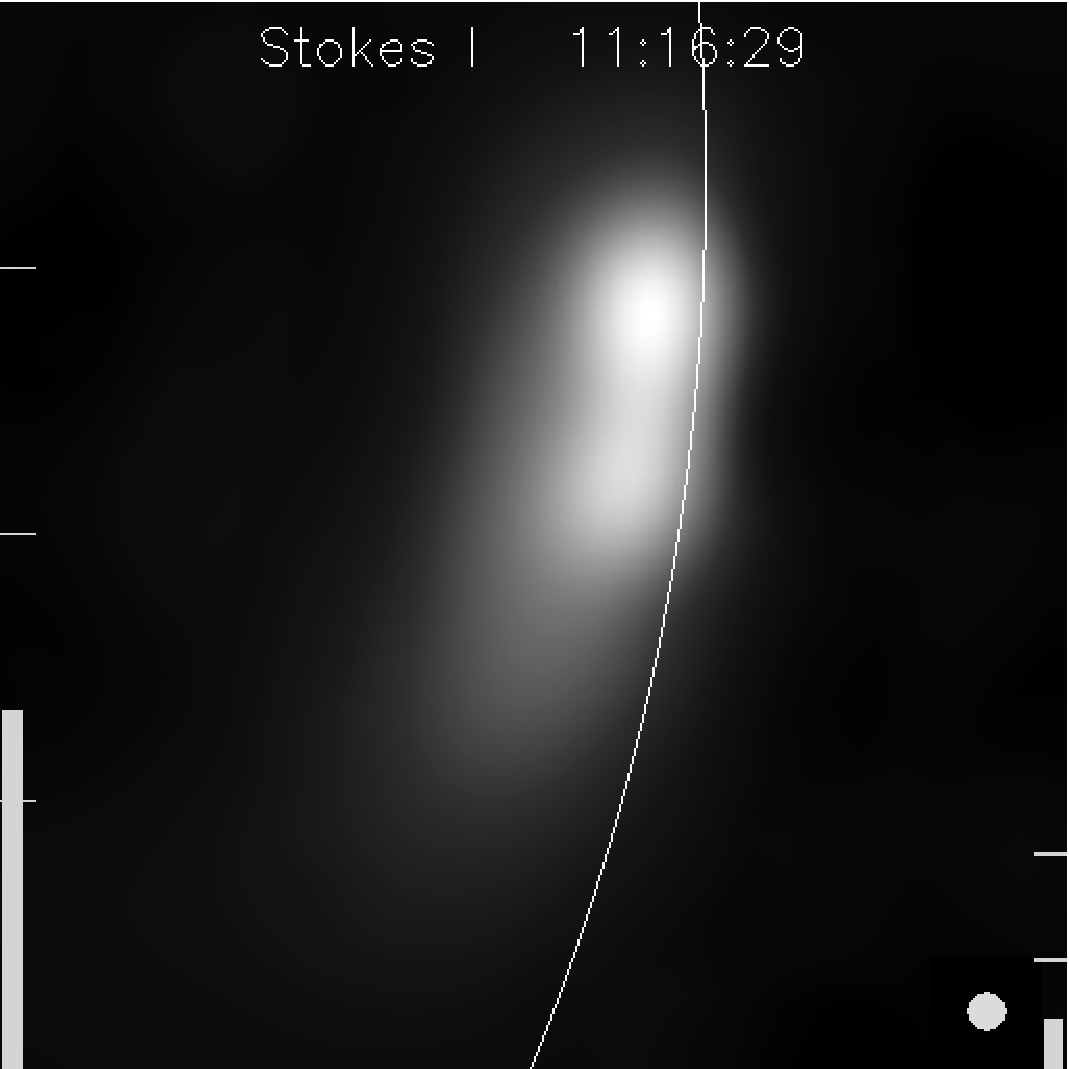}       % 7640 lambda
\includegraphics[width=4.45cm]{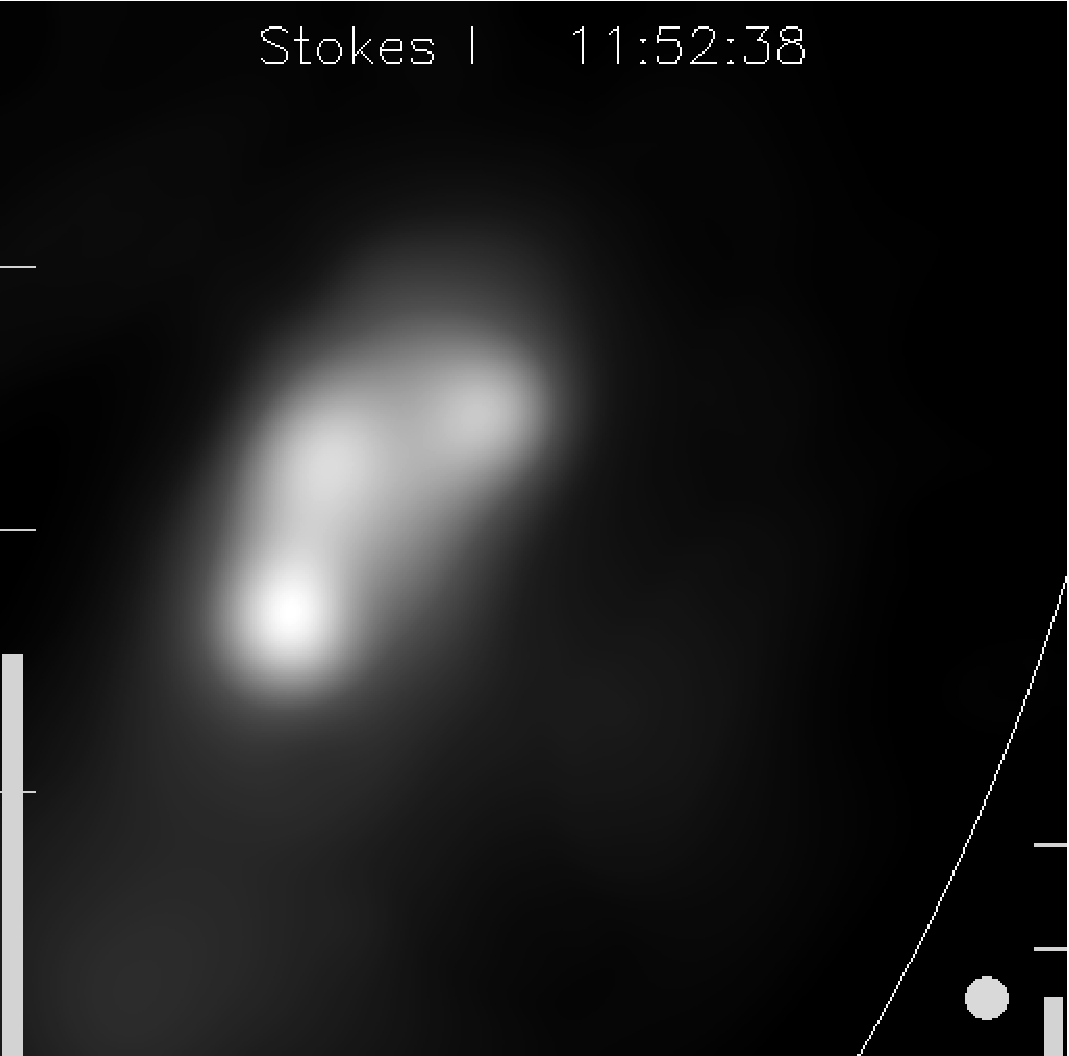} 
\caption{Composite images for Stokes I.   Left: two cores and a halo (all unpolarized) at 236 MHz on July 15, 2003 at 11:16:29.   The limits of the field of view (units of $R_{\odot}$) are ($0.70, 1.15)$ for EW and $(-0.38, + 0.07)$ for NS.  Right: three cores and a halo (LH polarized) at 327 MHz on Aug. 14, 2004, at 11:52:38.   The limits of the field of view (units of $R_{\odot}$) are ($0.50, 0.96)$ for EW and $(-0.50, -0.05)$ for NS.  The color scales are linear BW for both images.  The solid lines are the solar limbs.}
\label{images 2003jul15_2004aug14}       %  seul le 14 aug 2004 est referenc'e dans le texte.
\end{figure}
%\begin{figure}[!ht]     %   example 2004 aug 14   images  NB 7640 lambda  I et V refaite 01 avr 2014
%\centering
%\includegraphics[width=4.45cm]{2004aug14_327_11-52-38_I_f1.15.eps} 
%\caption{Example of composite images for Aug. 14, 2004 at 327 MHz for Stokes I at 11:52:38. Three cores are visible in the halo.}
%\label{image 2004aug14}
%\end{figure}
\begin{figure}[!ht]     %   example 2006 avr 06  refait 01 avr 2014
\centering
    \includegraphics[width=4.45cm]{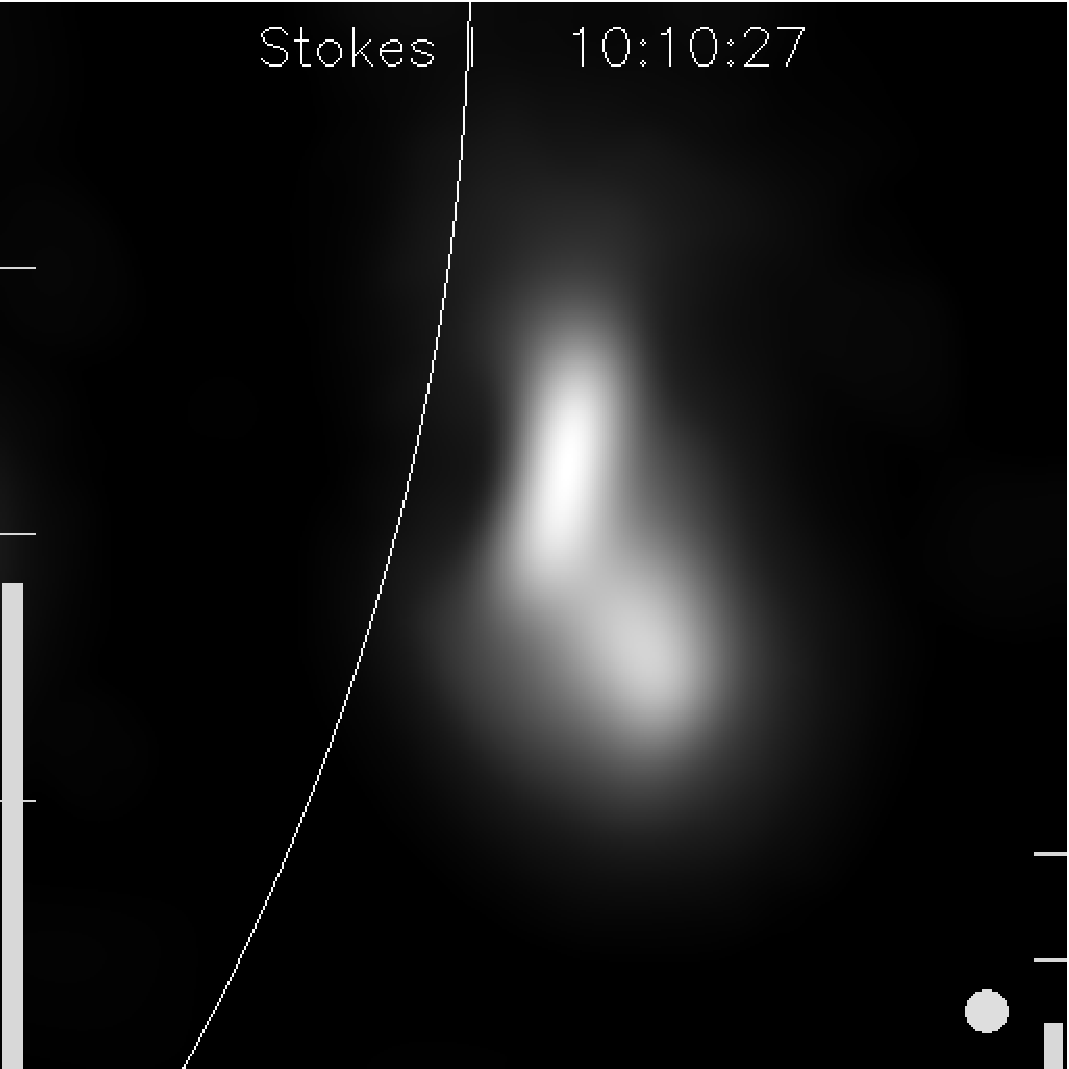}
    \includegraphics[width=4.45cm]{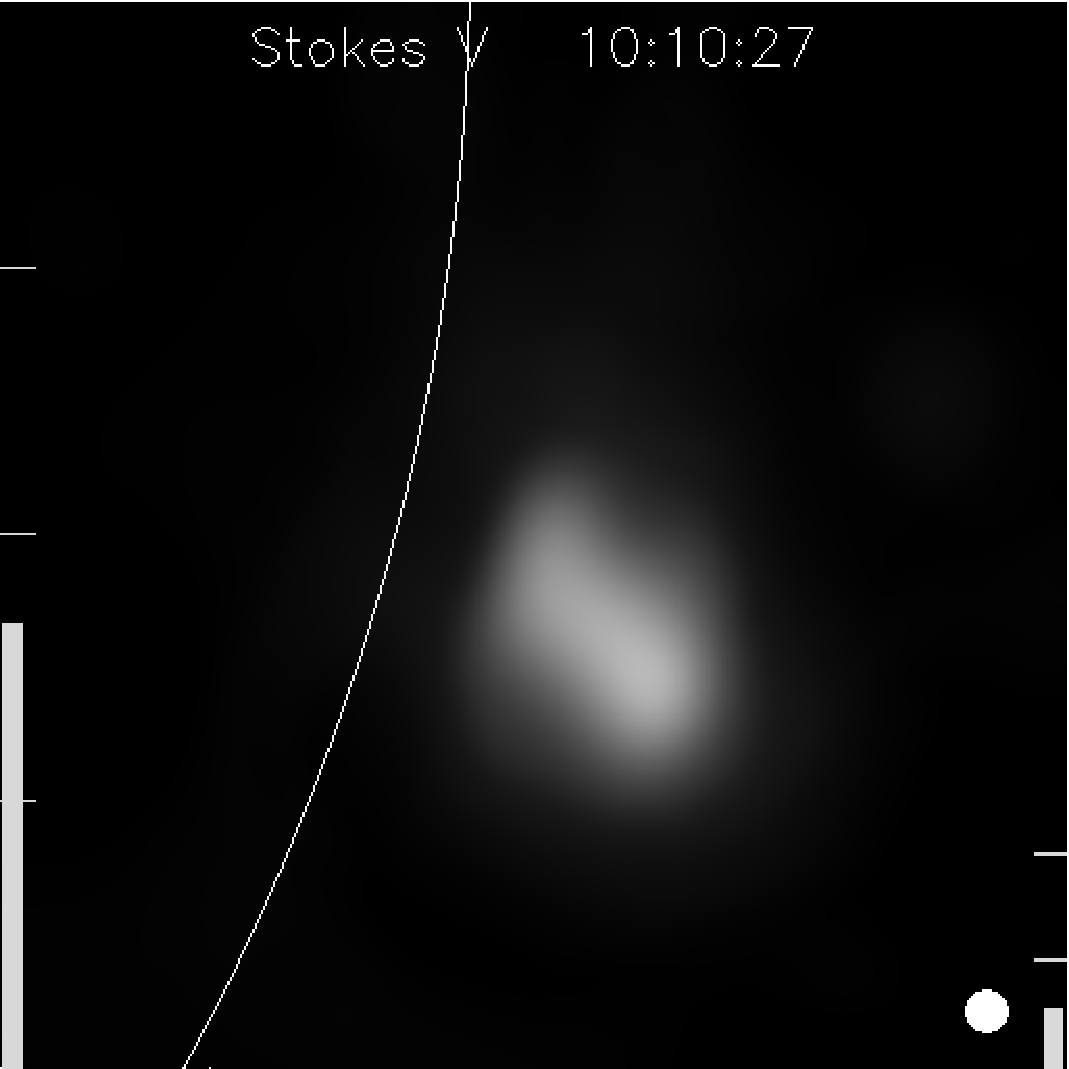}
\caption{Composite images for Apr. 06, 2006 at 236 MHz for Stokes I (left) and V (right) at 10:10:27.  The color scales are linear BW.  The limits of the field of view (units of $R_{\odot}$) are ($0.80, 1.25)$ for EW and $(-0.48, -0.03)$ for NS.  The solid line is the solar limb. The brightest core in I  is not seen in V.}
\label{images 2006avr06}
\end{figure}

\section{Results}
For three out of the four observations, there are several noise storms, one of them being more intense most of the time.  For these cases, we focused on the most intense storm.  As shown below, storms can be described as one or several cores with typical sizes 30-50 arcsec, embedded in a halo.  The polarization rate can change with position in the storm, but the sense of circular polarization remains the same.  The cores can be followed by continuity.  Their positions fluctuate by less than their sizes.  Their intensities vary in time by more than a factor 10, apparently in an uncorrelated way.  It can happen that one core becomes much more intense than the other.  The storm then appears simple and compact.  In this section, we give a qualitative description of storm structures from the GMRT / NRH images, and a quantitative analysis of their sizes.  We derive the altitude of storms at different frequencies using the NRH observations over several days.

\subsection {Qualitative description of the observed storms}
Examples of images are shown in Fig. \ref{images 2002aug27} to \ref{images 2006avr06}.  

On Aug. 27, 2002 (Fig. \ref{images 2002aug27}) there are three storms at 327 MHz.  The most intense ($T_b$ up to $10^9$ K) is near the western limb, is compact with a stable position during the observation, whereas its intensity fluctuates by a factor up to 8.  It consists of only one core.   It is strongly polarized, with the same aspect in Stokes I and V.  The weaker and more extended storms near the center of the sun have the same sense of polarization as the main. 

On Jul. 15, 2003 (Fig. \ref{images 2003jul15_2004aug14} left), there is only one unpolarized storm at 236 MHz, near the western limb.  This storm consists of an elongated halo with two cores, one being quasi-permanent at the northern end of the storm, the second (clearly separated) appearing only sporadically southward.  The positions of cores vary (apparently in a random way) by less than their widths.  The extent of the halo changes with time, extending toward the SE by up to 4 arcmin.  The brightness temperature $T_b$ is typically $2$-$3$ $\: 10^7$ K and sometimes reaches  $10^8$ K.  

On Aug. 14, 2004, three storms are visible at 327 MHz in the SW quadrant, separated by several arcmin.  The most intense storm (S1, shown in Fig. \ref{images 2003jul15_2004aug14} right) is always visible and elongated from SE to NW.  This storm consists of a halo with a banana-like shape and of several cores, all with LH polarization.  Four cores can be followed, with small changes in position.  Three of them are seen in (Fig. \ref{images 2003jul15_2004aug14} right) : C1 (at NW end) is the most intense one during most of the time with $T_b\sim 3 \: 10^7$ K , but C2 (at SE end) becomes briefly more intense around 11:48 UT,  with $T_b=7.4 \: 10^7$ K.   The polarization rates are 60\%  for C1 and 75\% for C2.  A fainter storm, southeast of S1, is always present, with RH polarization.  Another faint storm appears sporadically southwest of S1, with LH polarization. 

 % espaces :  \ ,   \ :    \ ;     (sans les blancs de lisibilite apres les \ )

On Apr. 06, 2006, the situation is still more complex.   Up to four storms are visible at 236 MHz.  The most intense storm (S1, shown in detail in Fig. \ref{images 2006avr06}) lies beyond the western limb,     with $T_b>10^8$ K and sometimes $>10^9$ K, and LH polarization.  Weaker storms can be seen when S1 is not too intense (Fig. \ref{aliasing} right) : S2, northeast of S1 and near the equator, is also LH polarized, S3 (east of S1) is unpolarized, and S4 (north of S1) is sporadic and unpolarized.
S1 consists of at least four cores, two or three of them often simultaneously visible (exceptionally 4).  In Fig. \ref{images 2006avr06} three cores C1, C2, and C3 (from SW to NE) are visible.  In Stokes I (left) C1 and C3 are the most intense ones.  C2 appears only as a southern extension of C3.  In Stokes V (right) only C1 and C2 are visible.  The position of cores may slightly change with time. Their relative intensity may vary by a factor $>10$ on  the integration time of the observation (2.1 sec.), apparently in a random way.  The brightness temperature $T_{b}$ may also change by a factor $\sim$ 10 on the same timescale.  The sense of polarization is the same over the whole storm.  The polarization rates range from $\sim80\%$ for C1 down to 1-2 \% for C3, and are constant for each core.  It follows that both shape and size of the storm change with time, and that they change differently for Stokes I and V : when a core (often C3) strongly dominates (for $\sim$ 10\% of the time), the storm size appears as minimum.  

In summary, for all cases but the first one, the storm structure is complex, with recognizable cores at slightly changing positions, embedded in a more extended halo. The sense of polarization is the same for the cores and the halo, but the polarization rate may substantially change among the cores.

\subsection {Apparent sizes of noise storms} 
From the descriptions given above, there are two spatial scales of interest in the structure of noise storms: the overall size, including cores and halo, and the size of individual cores.

\subsubsection {Overall size of noise storms}  
The shape of storms can be far from circular and the overall sizes were derived as follows.  At each time, we consider the N points with brightness between 0.49 and 0.51 $T_{b \: max}$, and the smallest rectangle with sides parallel to the main inertia axes of these points and which contains them.  The half-power widths along the main axes are defined as the dimensions of the rectangle.  For being robust, the procedure requires N $>$20.  If necessary, the image was interpolated to fulfill this condition.  The accuracy on the measured sizes then depends only on the quality of the image.  This quality can be appreciated from the level of the largest negative artifacts, which can be easily recognized as such in the cleaned images.  These artifacts are always few in number, their positions are stable and they obviously result from imperfect calibrations.  Their level, larger by more than one order of magnitude than the rms level of the background, is usually between 3\% and 15\%.  They cannot significantly affect the derived sizes.

%\begin{figure*}[!ht]		              %   largeurs  2002 aug 27  (faites 06 aug 2013)
%\centering
%\includegraphics[width=14cm]{2002aug27_12033_50_larg_pe.eps}
%\includegraphics[width=14cm]{2002aug27_12033_50_larg_gr.eps}
%\caption{Minor and major axes (top and bottom) at 327 MHz of the noise storm on Aug. 27, 2002 on composite GMRT + NRH images versus time (UT) : black is for Stokes I, red for Stokes V.  The maximum intensity is indicated in green (linear scale, arbitrary units).}
%\label {widths 2002aug27}
%\end{figure*}

%\begin{figure*}[!ht]		              %   largeurs  2003 jul 15   (faites 06 aug 2013)
%\centering
%\includegraphics[width=14cm]{2003jul15_7636_larg_pe_02sec.eps}
%\includegraphics[width=14cm]{2003jul15_7636_larg_gr_02sec.eps}
%\caption{Same as fig. \ref{widths 2002aug27}, for July 13, 2003 at 236 MHz.  Only sizes on Stokes I images are given, the storm being unpolarized.}
%\label {widths 2003jul15}
%\end{figure*}

\begin{figure}[!ht]    %   largeur petite  2004 aug 14   (faite 03 avr 2014)
                                    %   {figure*}  et  end {figure*}  pour utiliser 2 colonnes
\centering
\includegraphics[width=9.3 cm]{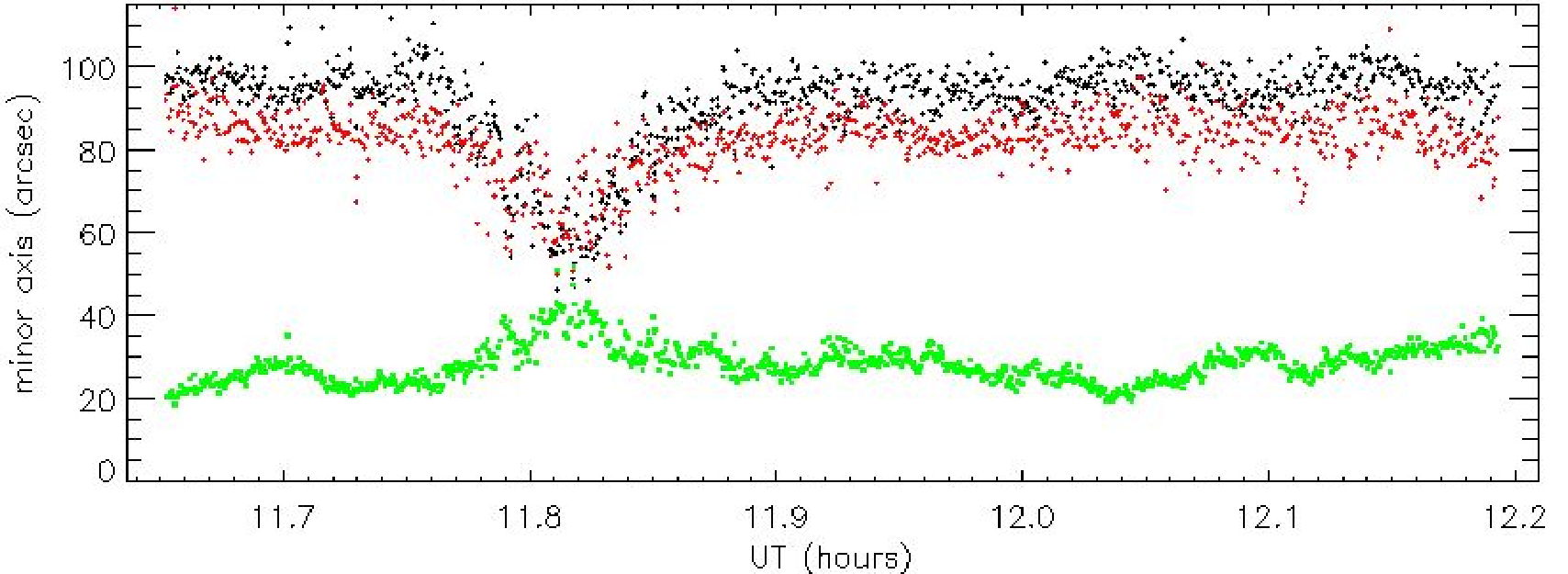}
% \includegraphics[width=9.3 cm]{2004aug14_327_larg_pe.eps}
           %  version 1 colonne  :  11 cm
           %  ----------  2 ----------s : 9.3 cm au max pour une seule colonne
\caption{Minor axis (arcsec) at 327 MHz of the noise storm on Aug. 14, 2004 on composite images versus time (hours, UT) : black is for Stokes I, red for Stokes V.  The maximum intensity is indicated in green (linear scale, arbitrary units).}
\label {width 2004aug14}
\end{figure}

%\begin{figure*}[!ht]		              %   largeurs  2006 avr 06   (faites 06 aug 2013)
%\centering
%\includegraphics[width=14cm]{2006avr06_8637_larg_pe.eps}
%\includegraphics[width=14cm]{2006avr06_8637_larg_gr.eps}
%\caption{Same as fig. \ref{widths 2002aug27}, for Apr. 06, 2006 at 236 MHz. Note the anti-correlation between sizes and intensity.}
%\label {widths 2006avr06}
%\end{figure*}

As an example, Fig. \ref {width 2004aug14} shows the variations of the minor axis for Aug. 14, 2004:  there is a marked minimum, down to $\sim 50$ arcsec around 11:48 UT, correlated with an increase of brightness temperature of the core C2.  The minor axis is always slightly larger in Stokes I than in Stokes V.  The dispersion is $\sim 5$ arcsec.  The large axis is about twice as large and has similar behavior.

On Aug. 27, 2002, the minor axis remains constant at $\sim$31  arcsec with very small fluctuations in spite of the intensity variations.  The major axis slowly increases from 40 to 60 arsec with some fluctuations, but the storm is never very elongated.  There are no significant differences between sizes in Stokes I and V.

On Jul.15, 2003, the minor and major axes are larger and scattered over 60-110 and 100-160 arcsec, respectively. This large scatter is consistent with the fact that the cores are never much brighter than the halo.  The scatter of both axes decreases with time while the intensity of the storm slowly increases.

%For Jul.15, 2003, the minor and major axes are larger and scattered over 60-110 and 100-160 arcsec respectively, consistently with the fact that the cores are never much brighter than the halo.  The scatter of both axes decreases with time while the intensity of the storm slowly increases.

On Apr. 06, 2006, both axes shows complex changes, because of the different time evolutions of the cores.  The minor axes are scattered over 35-100 arcsec, and the major axes over 70-140 arcsec.  Sizes on both axes are minimum for intensity peaks.  The smallest values of the minor axis ($\sim$ 35 arcsec) correspond to the intensity peaks of the core C3 (near the limb and weakly polarized).  As for Aug. 27, 2002, these values are close to those that are derived below for the cores themselves. Major axes are systematically smaller in Stokes V than in Stokes I. \\   

To summarize these results: 
\begin {itemize}
     \item    
       i) The minor and major axes have different behaviors on the four days. The major axis
         is more variable than the minor axis because of the presence of several cores, and it is smaller
         in Stokes V than in Stokes I because of the lower polarization degree of some cores.
    \item 
        ii) There is a anti-correlation between the variations of both axes and the intensity.  The 
          minimum sizes correspond to the dominance of one core and usually correspond to a large 
          $T_b$.  Conversely, the maximum sizes correspond to several cores being simultaneously 
          visible.  The two storms with the smallest sizes (Aug., 27, 2002, and Apr. 06, 2006) are also 
          the most intense.
    \item 
        iii)  In many cases, the overall sizes of storms at half level do not depend on the
          halo extent,  since the relative level of the halo is often below 50\%.  From a visual inspection 
          of the images, however, the total area where cores appear may be large part of the halo. 
          The  exception is July 15, 2003, for which the halo may sometimes extend far toward SE with 
          no clear presence of cores in this part of the halo.
\end {itemize}  
 
\subsubsection {Size of cores} 
\begin{figure}[!ht]	%   decroissance visib   2002 aug 27 Stokes I   et  2006 avr 06  Stokes IV
                                     % width=6.50cm  pour version referee:  Les : doivent etre remplaces par ; ou -
                                     %  et non par des  /  qui plantent a la compilation.
                                     % 
                                     % LE CHOIX DE LA LARGEUR DONNE DES RESULTATS CAPRICIEUX,
                                     %   avec des largeurs reelles differentes pour une meme valeur. 
\centering
\includegraphics [width = 4.40 cm ] {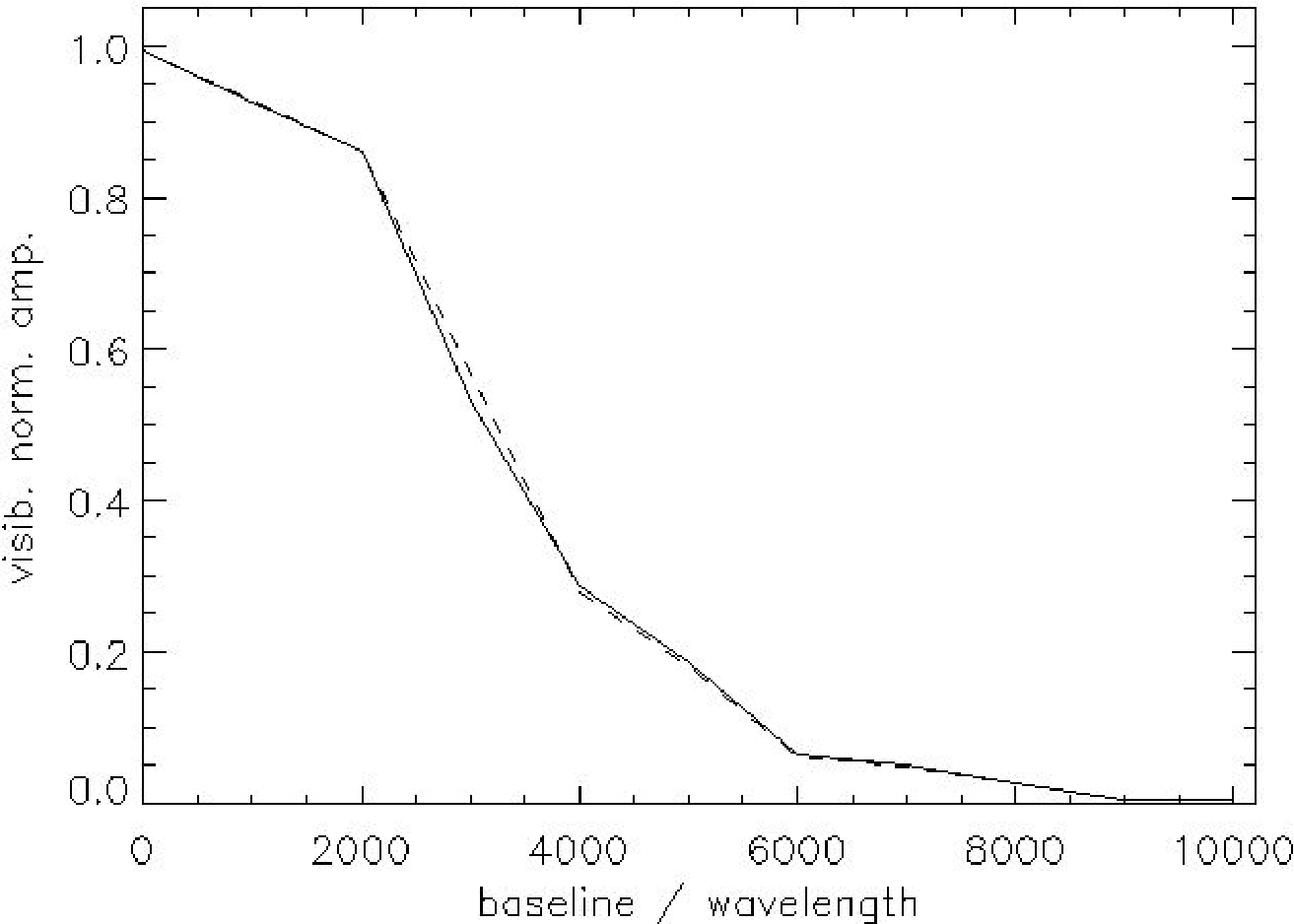}
\includegraphics [width = 4.40 cm ] {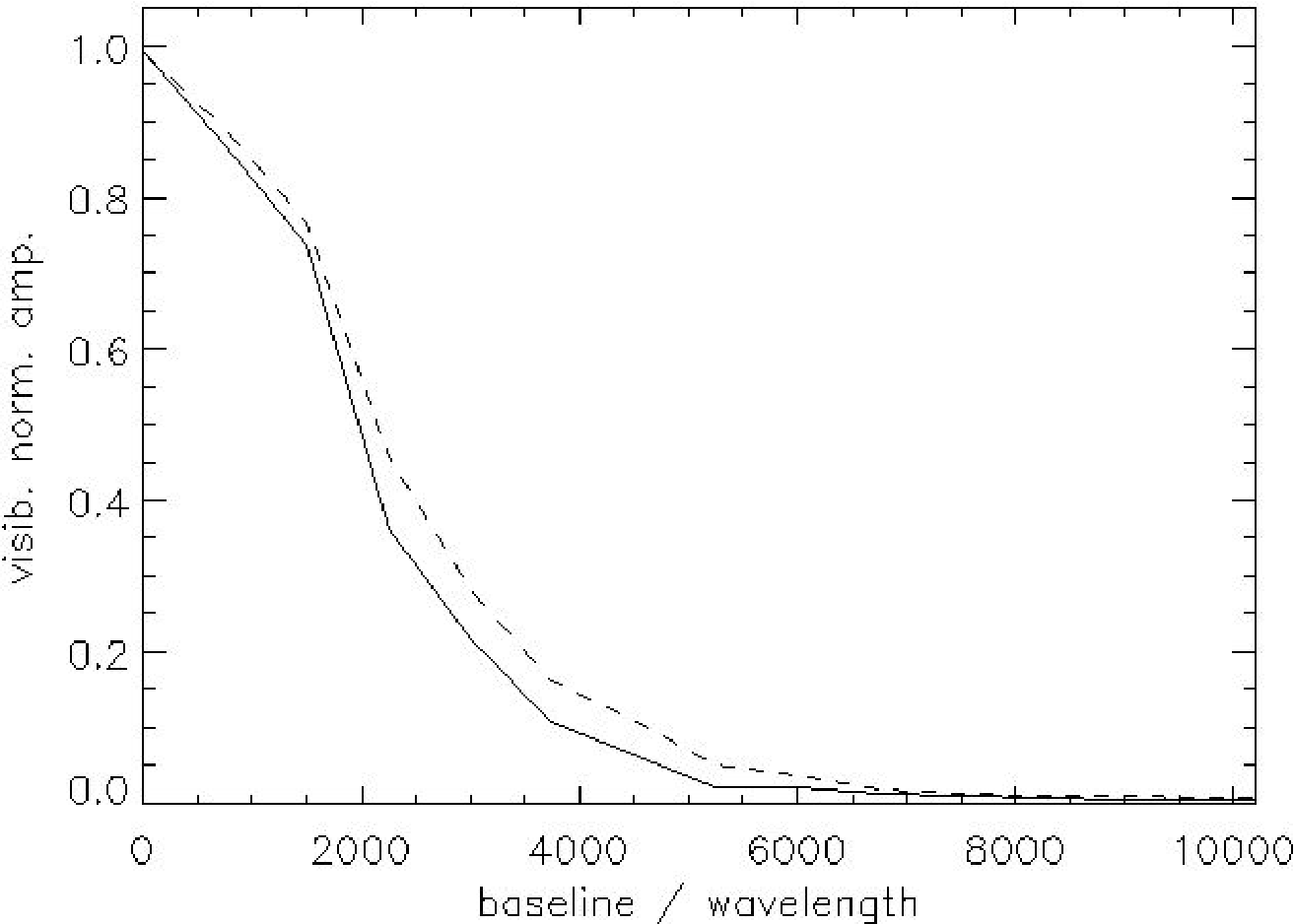}
% \includegraphics [width = 4.40 cm ] {decroiss_visib_2002aug27_09-15-05_IV.eps}
% \includegraphics [width = 4.40 cm ] {decroiss_visib_2006apr06_10-00-19_IV.eps}
    % version 1 colonne :   5.0 cm       version 2 colonnes  : 4.40 cm  au maximum
\caption {Normalized amplitude of Stokes I (solid line) and V (dashed line) azimuthally integrated complex visibilities versus the length of baselines (units of $\lambda$) for Aug. 27, 2002, at 09:15:05 UT (left) and for Apr. 06, 2006 at 10:00:19 UT (right).}
\label {decroiss visib 2002 2006}
\end {figure}

Core sizes can, in principle, be deduced from the extent of the complex visibility $V$ in the $uv$-plane, at least when the storm is dominated by an intense core.  The amplitude $\vert {V} \vert$ then decreases regularly to small values at distances $r_{uv} = \sqrt {u^2 + v^2} \: \sim 1/S$ from the origin of the $uv-$plane, $S$ being the size of the core.  In actual cases, $\vert {V} \vert$ decreases less regularly, according to the shape of the core itself and to the possible contribution of other storms or cores, but the method can give approximate values of the size of intense cores.  The extent of $\vert V \vert$ is limited to $\sim 2000$ $\lambda$ for Jul. 15, 2003 and Aug 14, 2004, but it is up to 6000 $\lambda$ or more for Apr. 06, 2006 and Aug. 27, 2002.  Fig. \ref{decroiss visib 2002 2006} shows examples for intense bursts on these two last days.  Table 2 summarizes the extent of the $\vert V \vert$ at relative levels of 0.5 and 0.1, and the sizes deduced under the hypothesis of Gaussian shapes.  These sizes are only indicative and are valid under the above assumptions.

%\begin {figure*}[!ht]    %   decroissance visib Stokes I et V  2004 aug 14 27  et  2006 avr  06
%\centering
%\includegraphics [width=6.7cm] {decroiss_visib_2004aug14_11;48;35_I.eps}
%\includegraphics [width=6.7cm] {decroiss_visib_2004aug14_11;48;35_V.eps}
%\includegraphics [width=6.7cm] {decroiss_visib_2006apr06_10;00;19_I.eps}
%\includegraphics [width=6.7cm] {decroiss_visib_2006apr06_10;00;19_V.eps}
%\caption {Same as for fig. \ref{decroiss visib 2002 2003} for Stokes I and V : Aug. 14, 2004 at 11:48:35 (top left and right) and Apr. 06, 2006 at 10:00:19 (bottom left and right).}
%\label {decroiss visib 2004 2006}
%\end {figure*}
\begin {table} [!h]  \begin {center}               % table des largeurs des visibilites et diametres deduits. 
\title {Table 2. Extents (rad$^{-1}$) at 50\% and 10\% levels of the amplitude of the complex visibility  for cases such as those shown in Fig. \ref{decroiss visib 2002 2006} , and derived half-power sizes (arcsec). } \\ [1.3ex]
\begin {tabular}{cccccccc} \hline\hline   % {llllr}  
                              &   at 50 \%  &   at 10\%   &    derived sizes  &  \\ %  [0.5ex]
%                              &                   &                    &         (arcsec)       &  \\ %  [0.5ex]
\hline
2002 Aug 27       &    2700   &   5900  &   31  &  \\
2003 Jul   15       &   1000    &   2000  &   87  &  \\  %  modifiees pour H>0
2004 Aug 14       &    900     &   2000  &   92   & \\  %
2006 Apr 06        &   2500    &   6000  &   33   & \\   % [1.0ex]
\hline                           % les tailles derivees ont ete verifiees le 22 avril 2014  (30 -> 31 pour 2002)
\end {tabular}
\end {center}
\end {table}
% Rappel sur les 1/2 largeurs a 1/2 et a 1/10 de puissance d'une gaussienne de largeur totale L :
%          u2 = 0.44 / L     u10 = 0.80 / L      ( calcul 16 mai 2013)
%       L          10"=4.848 E_5 rad       20"=9.696 E_5 rad     30"=1.345 E_4 rad      35"=1.697 E_4 rad
%       u2          9 076  lambda              4 540  lambda              3 025  lambda              2 593  lambda
%      u10       16 500  ---------               8 525  ----------               5 500  ----------              4 896  ----------
\begin {figure}[!ht]		              %   profils de noyaux
\centering
\includegraphics [width=4.40 cm ] {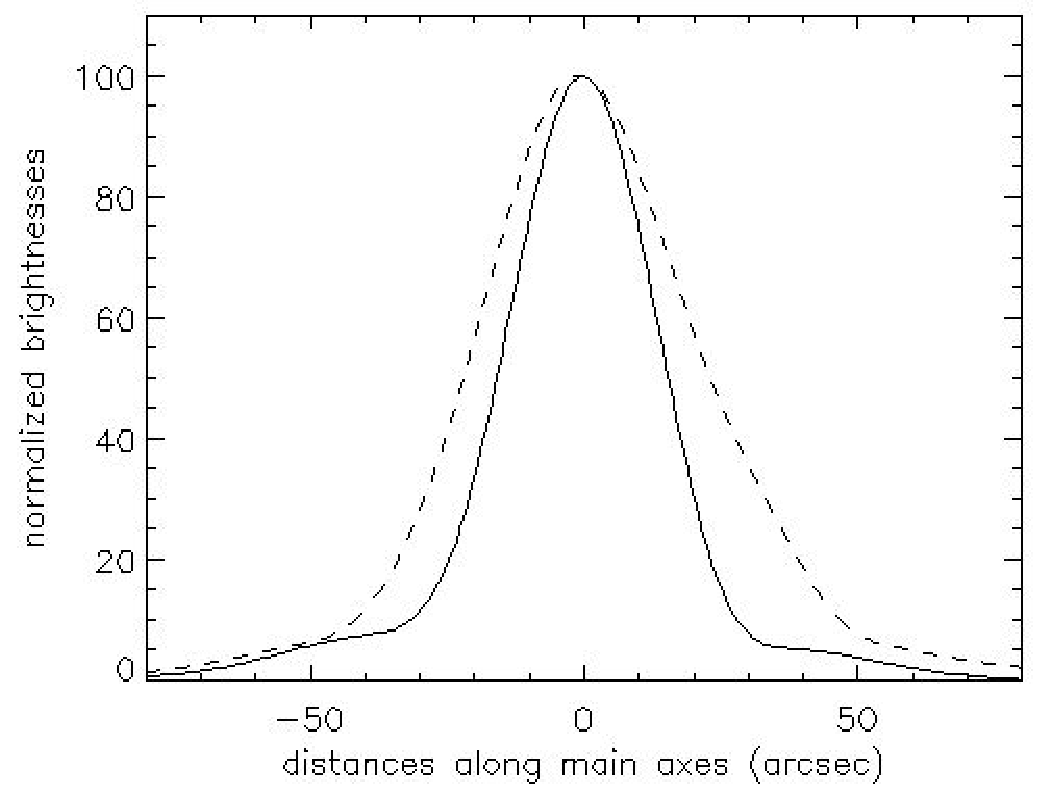}
\includegraphics [width=4.40 cm ] {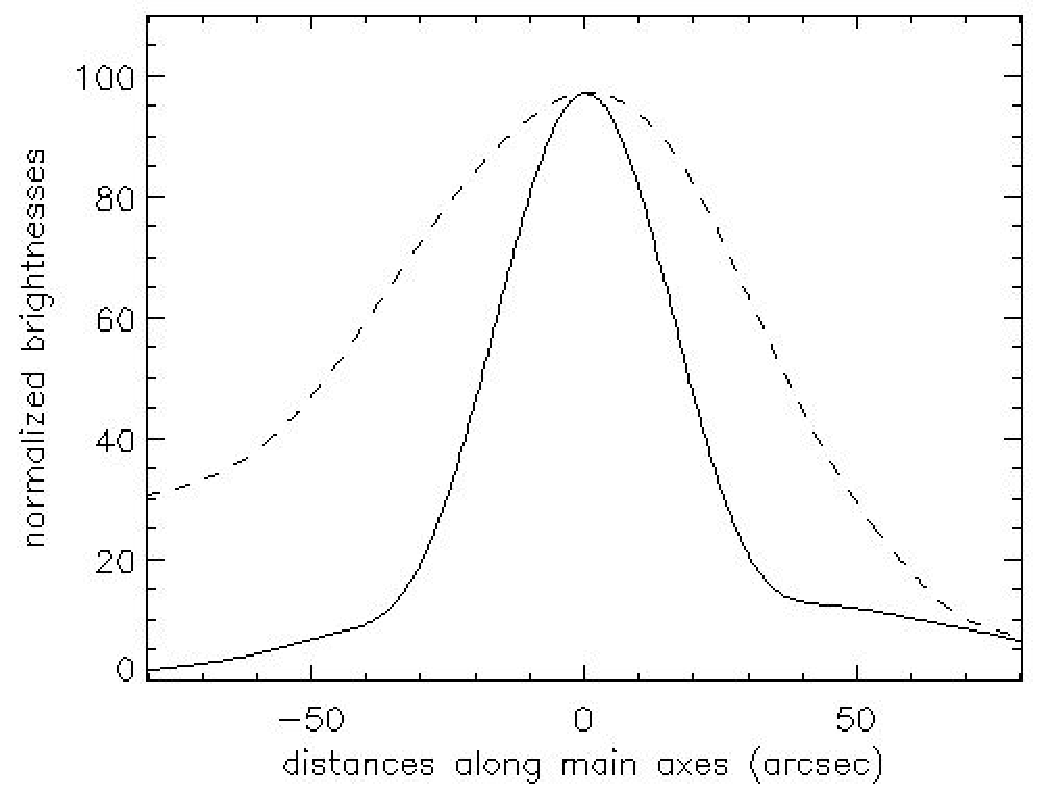}
% \includegraphics [width=4.40 cm ] {profils_I_2002aug27_09-01-34_+069_-034.eps}
% \includegraphics [width=4.40 cm ] {profils_I_2006avr06_10-42-28_+100_-013.eps}
    %  version 1 colonne   :  width=5.00 cm  cm,  height=4.10cm  
    %  version 2 colonnes :  width=4.45 cm  cm,  height=4.10cm  
\caption  {Examples of normalized profiles in Stokes I along minor and major axes (full and dashed lines) for intense cores on Aug. 27, 2002 at 09:01:34 UT (left) and Apr. 06, 2006 at 10:42:28 UT (right). Abscissas are in arcsec.}
\label {profils}
\end{figure}

The widths of cores can be directly measured from their 1D profiles when their intensity sufficiently exceeds that of other cores and of the halo, after applying approximate corrections for the contributions from the halo and from other cores.  Fig. \ref{profils} gives examples of profiles of cores among the narrowest cores along their main axes for Aug. 27, 2007 and Apr.06, 2006.  Similar profiles have also been obtained for Jul. 15, 2003 and Aug. 14, 2004.  Table 3 summarizes the results.  

Although the overall sizes of storms may substantially change with time or from one storm to another, the minimum sizes of cores listed in Table 3 are not very different, with the possible exception of Jul. 15, 2003, for which the intensity of the core is never much larger than that of the halo, rendering it difficult to measure its size.  
\begin {table} [!ht]  \begin {center}     % * pour prendre la largeur de la page. Il faut   \end {table*}
                                                                  %  Il faut  [!ht]  et non  [h], sinon la table est reportee a la fin;
\title {Table 3.  Smallest minor axes of cores for the four storms.}  \\ [1.5ex]     % [1.5ex] 
\begin{tabular}[c]{lccccccccc} \hline\hline   % {llllr}
date                      &    freq.   &     time )   &      width      \\
                              &  (MHz)  &     (UT)     &    (arcsec)   \\ 
\hline
2002 Aug.  27     &    327   & 09:01:34 &        31          \\   %[1.0ex]
2003 Jul.   15      &    236   & 11:04:57 &        57          \\   %[1.0ex]
2004 Aug. 14      &    327   & 11:49:04 &        45         \\ 
2006 Apr. 06       &    236   & 11:42:28 &        35         \\ 
%\hline\multicolumn {8} {l} {\small frequencies are in MHz and widths in arcsec} \\
\end{tabular}
\end {center}
\end {table}

\subsection{Heliocentric distances and vertical extents}
In the absence of stereoscopic observations, the heliocentric distances $r_{storm}$ at each frequency can only be derived by following the apparent position of noise storms during several days and assuming a rigid rotation.  This can only be done with the daily routine observations of the NRH at several frequencies.  Because of the limited NRH resolution, no distinction can be made between the positions of bursts and continuum.  It is known, however, that these positions may fluctuate from day to day, and even during one day (Mercier \emph {et al}, \cite {cons} at 164 MHz, Malik and Mercier, (\cite{malik} at 164, 236, 327 and 410 MHz).  This may introduce errors in the derived altitude, and only the self-consistency of the results gives us confidence in the derived altitudes.  

\begin{figure}[!ht]		              %   positions X   06 avr 2006  et echelle de hauteur derivee
\centering
\includegraphics[width=  4.45cm,  height = 3.4 cm] {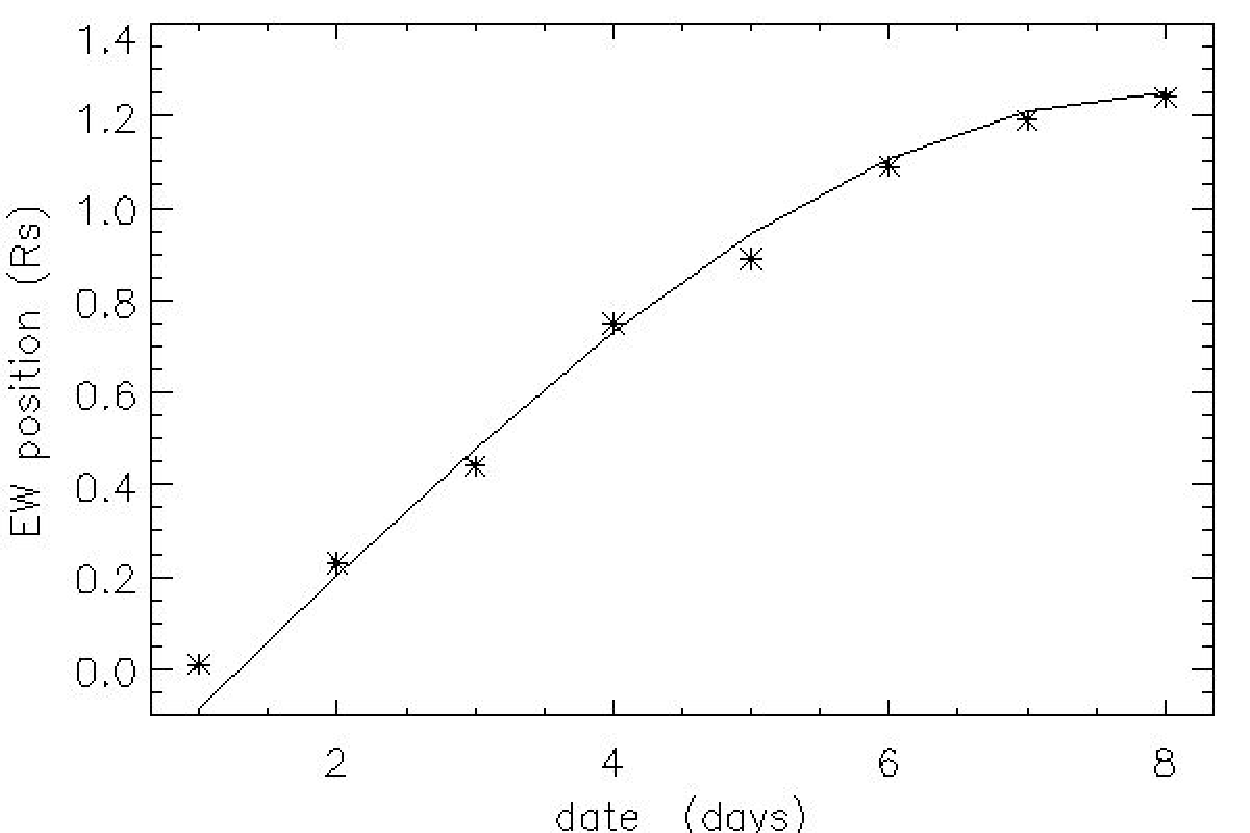}
\includegraphics[width = 4.45cm,  height = 3.4 cm] {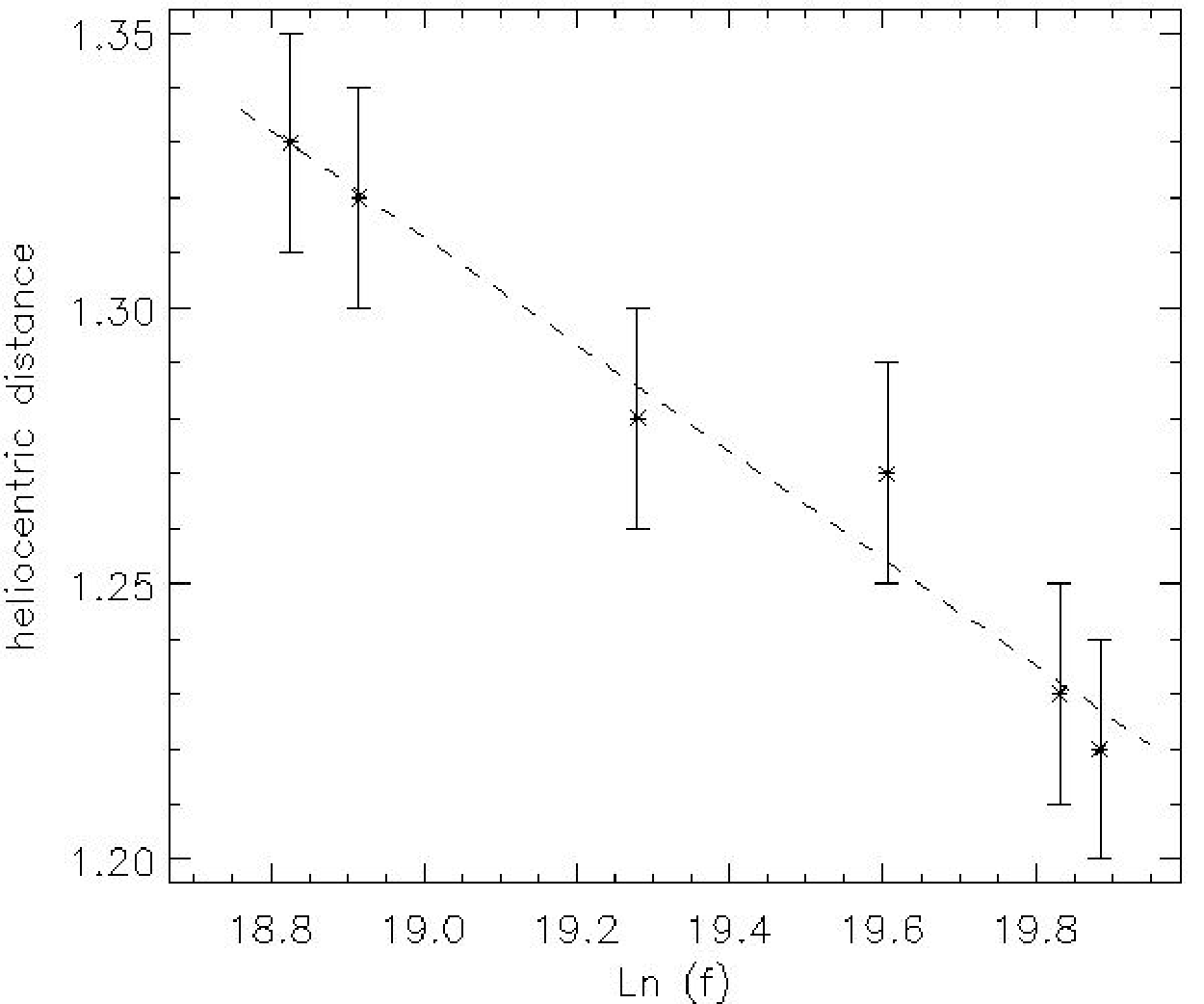}
% \includegraphics[width=  4.45cm,  height = 3.4 cm] {alt_2006_avr_06_327_X_B.eps}
% \includegraphics[width = 4.45cm,  height = 3.4 cm] {H0_2006_avr_06_B.eps}
     %  [width=5.0cm]    OK pour 1 fig par ligne
\caption{Left : apparent EW positions (units of $R_{\odot}$) of the main noise storm at 236 MHz from Apr. 1 to 8, 2006.  The solid curve is the fit with $r_{storm}=1.27 \: R_{\odot}$ and no refraction.  Right :  heliocentric distances (units of $R_{\odot}$), versus  Ln (frequency) for the same storm and linear fit (dashed line).}
\label{positions 2006 echelles}
\end{figure}

The prediction of apparent positions of rigidly rotating storms needs three parameters (time at the central meridian transit, heliocentric distance, and latitude) and the estimate of the shift due to refraction.  This shift vanishes at the center of the disk and is maximum at the limb, and it is smaller since emission arises in regions denser than the ambient corona.  It can be self-consistently neglected if the predicted positions can satisfactorily fit the observed positions during a whole transit on the disk.  As an example, Fig. \ref{positions 2006 echelles} (left) shows the apparent $EW$ positions at 327 MHz of the main storm on April 01-08, 2006 (there was no NRH observation on Apr. 09 and the storm was no longer visible on Apr. 10).  The fit with $r_{storm}=1.27$ $R_{\odot}$ and assuming no refraction is satisfactory for the whole period.  Fits are better at 432 and 327 MHz than at 150 MHz, storm positions being more stable at high frequencies. 

The derived $r_{storm}$ are listed in Table 4.  From the scatter of the data points in plots shown in Fig. \ref {positions 2006 echelles} (left), the errors can be estimated as $\sim$ 0.02 $R_{\odot}$.  In other words, if refraction effects were included in the model, they would not exceed this value very much.  Note that $r_{storm}$  almost systematically increases with decreasing frequency. 

%---------------------------------------------  pattern of the line length ----- -----------------------------------------------
 
\begin {table} [!h]  \begin {center}
\title {Table 4. Heliocentric distances (units of $R_{\odot}$) of noise storms vs frequency (MHz).} \\ [1.5ex]
%\title {Table 4. Heliocentric distances (units of $R_{\odot}$) of noise storms vs frequency.} \\ [1.5ex]
%
\begin {tabular}{lccccccc} \hline\hline   % {llllr}  
                              &   150  &   164   &    236  &   327   &   410   &    432   &  \\ %  [0.5ex]
\hline
% date \\
2002 Aug 27       &            &   1.25  &   1.25  &  1.20   &  1.18   &   1.20   & \\
2003 Jul   15       &   1.25 &   1.24  &   1.24  &  1.23   &  1.23   &   1.22   & \\  %  modifiees pour H>0
2004 Aug 14       &   1.30 &    1.27 &   1.24  &  1.21   &  1.18   &   1.19   & \\  %
% 2006 Apr 06    &  1.42  &   1.36  &   1.31  &  1.27   &  1.20   &   1.19   & \\   % [1.0ex]
2006 Apr 06        &   1.33  &   1.32  &   1.28  &  1.27   &  1.23   &   1.22   & \\   % [1.0ex]
\hline
%\multicolumn {8} {l} {\small heliocentric distances are in $R_{\odot}$.} \\
\end {tabular}
\end {center}
\end {table}

%---------------------------------------------  pattern of the line length ----- -----------------------------------------------

At all frequencies $r _{storm} \ge 1.2$ $R_{\odot}$, in agreement with previous studies (Le Squeren 1963, Elgar\o y 1977, Mercier and Kerdraon \cite{ker-mer-duino}).  The new point is that we have simultaneous \textit {2D} measurements at 5 or 6 frequencies, allowing us to derive the density scale height in noise storm sources.  Assuming that the emission frequency $f$ is equal to the local plasma frequency $f_p = 9\sqrt{n}$, where $n$ is the electron density (MKS units), and that the scale-height in the storm source 
                                      $H_{storm}=(-\frac{1}{n} \: \: \frac{dn}{dr})^{-1}$,
where $r$ is the heliocentric distance, is constant for emission between 150 and 450 MHz, we can write 
                         $ \: \: r=-2 \: H_{storm} \: Ln(f) \: + K \: \: $,  where $K$ is a constant.  
A least square fit of the data shown in Table 4 with this linear relation thus provides $H_{storm}$.  An example of fit is shown in Fig. \ref{positions 2006 echelles} (right) and resulting values of $H_{storm}$ are displayed in Table 5.  The typical uncertainty (0.02 $R_{\odot}$) in $r_{storm}$ at the various frequencies is not much smaller than the differences in height themselves, which results in high relative errors in the derived scale heights $H_{storm}$.  Using a standard procedure (Press \textit{et al}, \cite{press}), it can be estimated that $H_{storm}$ values are determined within $\pm10$ Mm.  In the case of July 15, 2003, the positions at different frequencies are so close to each other that it can only be said that $H_{storm}<30$ Mm.  

We introduce for convenience an overdensity factor $\:m \:$ as the ratio of the density in storms at heliocentric distance $r$ to that in quiet corona at the same $\: r\:$.  The latter is often estimated from the classical Newkirk's model (\cite{newkirk1}), which corresponds to an isothermal corona in hydrostatic equilibrium at a temperature $T_{cor}=1.4$ MK.  It explicitly refers to equatorial regions at cycle maximum and is the densest of all models.  The commonly cited Saito's equatorial model (\cite{saito}) is practically half as dense.  Chambe \& Mercier (\cite{cham-mer}) derived yearly averaged density models from measurements of the solar radio brightness beyond the limb for the period 2004-2011, both for equatorial and polar regions, in the same height range as that of noise storms.  Their equatorial models have scale heights similar to that of Newkirk's, but their densities are four times smaller on the average and decrease by a factor $\sim$ 2 between 2004 and 2009.   For our practical purpose, we have thus adopted a half dense model as Newkirk's, as an average model.   The values of  $\: m\: $  derived for each observing frequency are displayed in Table 5, in addition to the scale height  $H_{storm}$  and the scale height  $H_{cor}$  of Newkirk's model.
             
The $m$ factor is always substantially larger than unity and $H_{storm}$, in spite of its limited accuracy, is significantly less than $H_{cor}$.  The implications of  these high stom source densities on the propagation effects, and those resulting from the small scale-heights $H_{storm}$ on the emission models of storms are discussed in the following section. 

\begin {table*} [!ht]  \begin {center}
\title {Table 5. Density scale heights and overdensity $m$ factors in noise storms sources (see text).} \\ [1.5ex]   								%    [1.5ex]  regle l'interligne
\begin {tabular}{lcccccccccc} \hline\hline   % {llllr}  
% \multicolumn {5} {l} {\small  $<--$ T $= 1.5$ MK $-->$ $<--$ derived T $-->$} \\
                              &$H_{storm}$&  $H_{cor}$&   \multicolumn {6} {l} -$m \:$ values at each frequency (MHz) $\:$ - &  \\
                              &Mm           &       Mm     &  150  &  164   &  236  &  327  &  410  &  432  & \\
\hline
% Table rectifiee le 19 nov 2014, puis le 24 nov 2014 avec Newkirk1961 / 2
2002 Aug 27     &     30   ($\pm$ 10)  & 108     &            &    5.5    &    12  &   16  &   22   &   28  \\
2003 Jul  15      &       9   ($<20$)       &  111     &   4.6   &    5.2    &    11  &   19  &   30   &   32  \\
2004 Aug 14     &     36   ($\pm$ 10)  & 111     &   6.3    &   6.3    &    11  &    17  &   22   &  26   \\   
2006 Apr 06      &     35   ($\pm$ 10)  & 119     &   7.5   &    8.4    &    14   &   25  &   30   &  32   \\  
\hline
\end {tabular}
\end {center}
\end {table*}

%\end {document}

%---------------------------------------------  pattern of the line length ----- -----------------------------------------------

\section{Discussion}

\subsection {Internal structure of noise storms at a single frequency}
Kerdraon \emph {et al}. (\cite {ker-lang}) observed two bursts at different positions and times but without relating them to the overall storm structure. Lang and Willson (\cite {lang-will}), from data integrated over one hour, found evidence of fine structure involving several compact sources embedded in a more extended source.  They gave snapshot images (integration time13 and 30 s) from the same observation, showing bursts occurring at different positions, but they gave no relation between these positions and that of storm features visible in the time-integrated image.  

Our results confirm and extend this description of storms.  We find that they consist of a diffuse region, referred to as the halo, the extent of which is $\sim$ 2 arcmin and may change with time, and one or several more compact and brighter cores, embedded in the halo.  
The halo does not necessarily exist in all cases and its presence cannot be ascribed to instrumental effects. $\,$  For instance, combining complex visibilities that are densely sampled but have a limited extent around the origin (from the NRH), with a more extended and sparser set of complex visibilities (from the GMRT) produces an inhomogeneous $uv$-coverage which could result in a composite beam with a core and a halo-like structure.   However, the effect of the deconvolution procedure is to correct the inhomogeneities in the $uv$-coverage density. $\,$ Moreover, we note that : $\,$ i) on Aug. 27 (Fig. \ref{images 2002aug27}), there is no halo (or very weak) around the main storm near the western limb, whereas the source near the center of the disk is complex with a time-varying halo, $\,$ ii) on Aug. 14, 2004, the positions of the halo and of the cores are stable, but the most intense core is not always the same (often at north, then at south at the maximum intensity at 11:48 UT), and $\,$ iii) on Jul. 15, 2003, conversely, the extent of the elongated halo can change within a few seconds, with no changes in the position of cores.  These changes cannot arise from instrumental effects and we must conclude that the observed core$/$halo structure is physical. 
 
Both halos and cores can be visible on most of the snapshot images.  Although the cores may be recognized from image to image, their position may fluctuate.  The changes in their relative intensities seem stochastic and can be large, so that only some of them can be visible at a given time.  At intensity peaks, the image is generally dominated by one bright core.  Although our observation rate (2sec) is too slow to distinguish between continuum and bursts, bursts themselves should originate from cores.  The sense of circular polarization is the same over the whole extent of the storm, but the polarization rate may strongly differ from one core to another. 

\subsection {Apparent sizes of noise storm cores}
At 327 MHz, Lang and Willson (\cite {lang-will}) found core sizes of $\sim$ 40 arcsec in an image averaged over a period of 1h where the storm level was stable (hence presumably for the continuum) as well as in a few snapshot images at times of intensity peaks (presumably for Type I bursts).  Zlobec \emph {et al} (\cite {zlobec}) reported sizes $\sim$ 50 arcsec for about 20 bursts, with a minimum reliable value of 40 arcsec for two of them.   Kerdraon \emph {et al} (\cite {ker-lang}) also reported two bursts with size of 40 arsec.       In our study, the main noise storm on Aug. 27, 2002 has  a nearly stable size in the 224 images during the whole observation of 1 hour, with a minimum of 31 arcsec.  At 236 MHz, no sizes were previously reported.  We find a minimum size of 35 arcsec for some intense bursts from the core C3 on Apr. 06, 2006.  On  Jul. 15, 2003, the minimum value of 57 arcsec is less reliable since the corresponding core was never much brighter than the halo.

\subsection {Coronal density and scale height in noise storm sources}  
We recall briefly the results of previous studies on the altitude of radio noise storms and/or their relative positions at different frequencies.  At 408 MHz, Clavelier (\cite{clavette}), using the former EW 1D Nan\c{c}ay interferometer, found a mean $r_{storm} \sim1.1 R_{\odot}$.  At 200 MHz, Morimoto and Kai (\cite{morimoto-kai}), using the 1D Tokyo interferometer, derived the range 1.2-1.3 $R_{\odot}$. At 169 MHz, Le Squeren (\cite{lesqueren}), from an extensive study with early versions of Nan\c{c}ay EW and NS interferometers, derived $r_{storm}$ between 1.2 and 1.9 $R_{\odot}$, with a mean value 1.5 $R_{\odot}$. Using the NRH in its early 2$\times$1D version at 169 MHz, Kerdraon and Mercier (\cite{ker-mer-duino}), obtained the mean value 1.22 $R_{\odot}$.  

These studies are heterogeneous since most of them used observations done with various instrumental limitations (e.g., only one-dimensional resolution, limited position accuracy, single observing frequency), and at different stages of the solar cycle. In spite of this, a general trend is that noise storms lie relatively high in the corona, involving larger densities than given by usual coronal models.  It was, however, not possible to draw a conclusion concerning the scale height in noise storm sources: the results of Le Squeren (\cite {lesqueren}) and Clavelier (\cite {clavette}) seem to indicate a large difference in altitudes at 169 and 408 MHz, but their observations were not simultaneous and were carried out with different first-generation instruments.  Moreover, the high altitudes found by Le Squeren (\cite {lesqueren}) are not consistent with those found by other authors. 

The first  multifrequency observations with the NRH at 164, 236, 327, and 410 MHz were reported by Malik and Mercier (\cite{malik}).  They found  that the apparent positions of the continuum at different frequencies were often closer to each other than 0.06 $R_{\odot}$ and had strongly correlated small-scale motions.  However, they did not derive heliocentric distances and consequently gave no explicit results on the overdensity of noise storms and on their density scale height.   

We obtain, for the first time, heliocentric distances for the same noise storm at several frequencies, using NRH observations over several days.  These heliocentric distances are in the same range as those found in earlier studies, but the new point is that they are close to each other at the different frequencies.  This is consistent with the small differences in position with the frequency found by Malik and Mercier (\cite{malik}).  The values of the overdensity factor $m$ listed in Table 5 are relative to a  model similar to that of Saito (\cite{saito}).   Using models from Chambe $\&$ Mercier (\cite{cham-mer}) would yield still higher $m$ values.

These $m$ values would be reduced if storm sources were located in or near active regions (AR).  The comparison with images from EIT aboard SOHO shows that for all cases but Jul. 15, 2003, the noise storms were near the border of active regions (AR), yet clearly outside.  For Jul. 15, 2003,  the storm was between two AR.  Hence for all cases, storms are not located in high density regions and the $m$ factor should remain large.  This has consequences for the scattering effect, as discussed below. 

In spite of their limited accuracy, the density scale heights in storm sources $H_{storm}$ are found to be smaller than in the ambient corona.  Two arguments can be given to ensure that this result is not biased by refraction.  The first involves an estimate of the refractive effects needed to account for the apparent position differences between $f_1$ = 432 MHz and $f_2$ = 150 MHz.  If the actual scale height in storm sources was hydrostatic, as in the ambient corona, the difference $r_2 - r_1$ between heliocentric distances of the plasma levels $f_1$ and $f_2$ could be expressed as :  
%\begin {equation}
$\;  r_2 - r_1 = r_1 \big ( \frac{1}{1 - A \: r_1} - 1 \big ) $   %  \quad ou qquad  => grand espace.
%\end {equation}
where $r_1$ and $r_2$ are in units of  $R_{\odot}$ and $A=\frac  {2 k_B \: T}   {M \: g_0 \: R_{\odot}} Ln(\frac{f_1}{f_2}) $.  Taking $T$=1.5 MK and $r_1$ = 1.2 gives $r_2 - r_1$ = 0.48, whereas $\sim$ 0.1 is observed.  A differential shift of 0.38 $R_{\odot}$ would then be needed to account for the observed relative positions.  Apparent shifts for the positions of sources at $f_1$ and $f_2$ would be still larger, in contradiction with the fact that the maximum refraction effects cannot much exceed $\sim$ 0.02 $R_{\odot}$.  The second argument comes from Type III bursts, which are produced by fast electron streams accelerated at the borders of AR and propagating along field lines.  Their circular polarization rate is low and it is accepted that the emission takes place at $2 \: f_p$.  Their apparent altitude at the limb in the NRH frequency range is similar to that of noise storms.  Thus, if the small position differences with frequency for noise storm was due to refraction, the same would be expected for Type III bursts.  Although no systematic study has been done, NRH data usually show larger position differences with frequency than for noise storms.  Examples are given by Klein \textit{et al} (\cite{klein-2}).  We conclude that the small values for $H_{storm}$ are not due to refraction effects, consistent with the fact that these effects must be small because of the high altitude of noise storms. 

The fact that $H_{storm} < H_{cor}$ can be explained in two ways: either the temperature is very low in the source, $\sim$ 0.6 MK, which seems unlikely for a region where suprathermal particles are produced through magnetic dissipation, or $H_{storm}$ does not correspond to an hydrostatic equilibrium.  Since such an equilibrium should be effective along magnetic field lines, this last possibility implies that regions emitting at different frequencies do not lie on the same field line.  This leads to the following possible sketch for noise storm sources : overdense and compact regions, extending normally to the magnetic field with strong density gradients normal to the magnetic field. In at least three out of the four cases reported here, the density decreases upward.  This is consistent with our visual experience of NRH data, showing that for most noise storms near the limb, low- frequency sources appear slightly farther from the center of the disk than those at high frequencies.  

% espaces :  \ ,   \ :    \ ;      (sans blancs de lisibilite apres les \)    \quad ou qquad  => grand espace.

\subsection {Implications on the columnar model and the emissions theories}
Lang and Willson (\cite {lang-will}) had already pointed out that the complex structure they observed in a noise storm was difficult to reconcile with the simple geometry of the columnar model.  We confirm this complexity and describe it in more detail.  We also add another argument against the columnar model: the density scale height $H_{storm}$ in storm sources, derived from multifrequency observations, is too small for storm sources at different frequencies lying along the same magnetic field line, as explained above.  This questions the theoretical emission models, which require trapping of supra-thermal electrons in a closed magnetic loop, from which the columnar model naturally follows.  The high polarization rate of storms is usually ascribed to the fact that radio waves are emitted close to the local plasma frequency, and that, because of the magnetic splitting of the plasma cut-off, only the ordinary mode can escape.  It is thus difficult to understand how different polarization rates, and particularly very low, can arise from close sites. 

As things stand, multifrequency and high-resolution observations would be useful for better constraining models of noise storms structure and of emission mechanisms.  In particular, it is currently not known how the positions of cores at different frequencies relate to each other.  

\subsection {Implication on scattering effects}
The apparent size of noise storms is currently explained by propagation effects in a turbulent corona.  Bastian (\cite{bastian}) and  Subramanian \& Cairns (\cite{subra-cairns}) have modeled these effects assuming small angle scattering for radiation from compact sources embedded in the corona.  The limit of validity of this assumption has been estimated by Bastian to a broadening of 25 arsec.  This is not much smaller than the smallest sizes reported here (31 and 35 arc sec at 327 and 236 MHz, respectively), which can be partly real.   These authors adopt different models of density fluctuations in the corona, and predict widely different broadening of an ideal point source.  At 300 MHz, Bastian predicts apparent sizes of at least several tens of arcsec, well outside the validity domain.  In contrast, Subramanian and Cairns, using a lower turbulence level, predict sizes of less than one arcsec. Our observations, at the validity limit of the models, show that the actual turbulence level, which is presently  still poorly known, should be intermediate and they could help to specify it.

The constraints on scatter broadening at 327 MHz from our study are only marginally stronger than from the previous studies, since the minimum size we find (31 arcsec) is not much smaller than those previously obtained.   However, at 236 MHz, the smallest size we observe (35 arcsec) is hardly larger than that at 327 MHz, whereas broadening through scattering is expected to scale as $1/f^2$.  In principle, this scaling holds for refractive index $\mu$ close to unity but Subramanian and Cairns showed that taking into account the departure of $\mu$ from unity only brings unimportant changes in the broadening.   In any case, storm sources are compact and overdense and the frequency of the emitted radiation is well above the local plasma frequency as soon as it escapes from the source.  The fact that we do not find the expected ratio can be explained in two ways: either the apparent sizes are essentially real, or the turbulence level changes with time and space.  In this last case, the turbulence level should have been more or less the same for the five storms observed at 327 MHz (including ours)  and lower for the two storms we observed at 236 MHz, particularly for Apr. 06, 2006. 

More definite conclusions would require future simultaneous high-resolution imaging at several frequencies (preferably more than two) of several storms, to separate scattering effects and actual changes in size with frequency.  This is not possible however with current radiotelescopes.  All the storms we reported here were near the limb and that it could be helpful to obtain further observations such as ours for storms near the center of the disk, where scattering effects are smaller.  

\section{Conclusion}
We have shown that combining visibilities from the NRH and GMRT works well and is useful, providing snapshot  images with a high dynamical range, a wide field of view, and a high spatial resolution.  These characteristics were essential in the present study since noise storms show internal structure and since several storms often coexist.  Even with the few cases studied here, we get new insights on the structure of noise storms. 

It was already known that the electron density in noise storm sources exceeds that in the ambient corona.  We specified overdensity factors of 5-25 relatively to the widely accepted Saito's  model, and even more relatively to quiet corona models derived from purely radio observations.  

From multifrequency NRH observations, we derived the scale height of the electron density in noise storm sources and showed that it is smaller than in the ambient corona.  This implies that the coronal regions emitting at different frequencies do not lie along the same magnetic flux tube.  This questions the classical columnar model and also the current theories  for emission mechanism, which imply magnetic trapping of suprathermal electrons of a few keV.  

Noise storms appear to have an internal fine structure with one or several bright and compact cores embedded in a more extended halo.  The positions of cores fluctuates by less than their size over a few seconds.  Their relative intensities may change over time of 2 s, implying that bursts originate from cores.  It follows that the overall apparent shape and size of storms may change rapidly, giving the impression of being quasi-random.  The sense of circular polarization is the same over the whole storm.  The polarization rate is stable for each core, but may differ between the cores: for Apr. 06, 2006 it is $\sim$ 80 \% for the two cores at the southern end of the main storm and practically 0 \% for the more intense and elongated core (C3) lying just above (Fig. \ref {images 2006avr06} left).     

The minimum observed sizes of cores are of interest for discussing scatter broadening.  At 327 MHz, we observed a compact storm with a remarkably stable size during the whole observation (1hour), with a minimum value of 31 arcsec, slightly smaller than those previously reported (40 arcsec).    At 236 MHz, the smallest sizes we found (35 arcsec) correspond to the highest intensities of a particular core in a complex storm.  It is presently difficult to conclude whether these apparent sizes are real or broadened by scattering, considering that the predictions of current theories are limited by the poor present knowledge of the turbulence level and of its space and time variations.  In addition, there are too few reliable observations with high spatial resolution.  More observations of storms at various solar longitudes could be helpful.  However, conclusive observations of storms at several simultaneous frequencies with high spatial resolution ($<10$ arcsec) and time resolution ($<$ 1sec), in order to observe the same storm at different levels and to clearly separate bursts and continuum, does not appear appear feasible with currently operating instruments.  

%\end {document} 

%---------------------------------------------  pattern of the line length ----- -----------------------------------------------
 
\begin {acknowledgements}
We thank Stephen White who edited the GMRT data of April 06, 2006 when he was a visiting scientist in Meudon in July 2006.  P. Subramanian acknowledges partial support from the CAWES-II program, administrated by the Indian Space Research Organization.  We also thank N. G. Kantharia and S. Ananthakrishnan for helpful discussions regarding GMRT observation procedures, and an anonymous referee for helping us improve the manuscript.
\end {acknowledgements}

%---------------------------------------------  pattern of the line length ----- -----------------------------------------------

\end{document}